{}%disable hyper at arxiv
{}%use pdflatex at arxiv

\documentclass[12pt,a4paper]{article}

\newif\ifpublic\publicfalse
\newif\ifniklas\niklastrue
\newif\ifniklas\niklastrue

\usepackage{ifpdf}
\ifniklas\ifpdf\publictrue\else\publicfalse
\PassOptionsToPackage{hypertex}{hyperref}
\PassOptionsToPackage{draft}{graphicx}
\fi\fi

\usepackage[utf8]{inputenc}
 \usepackage[T1]{fontenc}
\usepackage{amsmath,amssymb,amsthm}
\ifpublic\else\usepackage{showkeys}\fi
\usepackage[bookmarks=true,hyperfigures=true]{hyperref}
\usepackage{graphicx}
\usepackage{dsfont}
\usepackage{xcolor}
\usepackage[nosort]{cite}
\usepackage[bulletsep]{collref}
\usepackage{bm}
\usepackage{multirow}
\ifniklas

\def\showkeysrefformat#1{{\normalfont\tiny\ttfamily#1}}
\makeatletter
\def\SK@@ref#1>#2\SK@{%
 {\@inlabelfalse\leavevmode\vbox to\z@{%
 \vss\SK@refcolor\rlap{\vrule\raise .75em%
  \hbox{\showkeysrefformat{#2}}}}}}
\makeatother
\fi

\usepackage[a4paper,text={160mm,247mm},centering]{geometry}
\allowdisplaybreaks[3]

\numberwithin{equation}{section}

\usepackage[font=small,labelfont=bf,width=0.85\textwidth]{caption}

\expandafter\def\expandafter\bfseries\expandafter{\bfseries\ifmmode\else\boldmath\fi}
\expandafter\def\expandafter\mdseries\expandafter{\mdseries\ifmmode\else\unboldmath\fi}
\expandafter\def\expandafter\normalfont\expandafter{\normalfont\ifmmode\else\unboldmath\fi}

\RequirePackage{verbatim}

\makeatletter
\newwrite\bibinl@out
\newenvironment{bibtex}[1][\jobname]{%
  \immediate\openout\bibinl@out #1.bib
  \immediate\write\bibinl@out{\@percentchar generated from `\jobname' starting line \the\inputlineno^^J}%
  \def\verbatim@processline{\immediate\write\bibinl@out{\the\verbatim@line}}%
  \@bsphack\let\do\@makeother\dospecials\catcode`\^^M\active\verbatim@start
}%
{\immediate\closeout\bibinl@out\@esphack}
\makeatother

\usepackage{fixmath}

\newcommand{\sfrac}[2]{{\textstyle\frac{#1}{#2}}}
\newcommand{\half}{\sfrac{1}{2}}
\newcommand{\ihalf}{\sfrac{i}{2}}

\newcommand{\alg}[1]{\mathfrak{#1}}

\newcommand{\mcA}{\mathcal{A}}

\newcommand{\lH}{\mathcal{H}}
\newcommand{\lQ}{\mathcal{Q}}
\newcommand{\lB}{\mathcal{B}}

\newcommand{\lK}{\mathcal{K}}

\newcommand{\bH}{\mathbb{H}}
\newcommand{\bQ}{\mathbb{Q}}

\newcommand{\su}{\mathfrak{su}}

\newcommand{\co}{\mathbf{c}}
\newcommand{\no}{\mathbf{n}}

\newcommand{\cdo}{\mathbf{c}^\dagger}

\newcommand{\vac}{|0\rangle}

\RequirePackage{amsmath}
\makeatletter
\newcommand{\brk@ord}{\bBigg@{0}}
\newcommand{\brk@ordl}{\mathopen\brk@ord}
\newcommand{\brk@ordr}{\mathclose\brk@ord}
\newcommand{\brk@ordm}{\mathrel\brk@ord}
\newcommand{\brk@var}{\brk@ord}
\newcommand{\brk@varl}{\left}
\newcommand{\brk@varr}{\right}
\newcommand{\brk@varm}{\mathrel\brk@var}
\newcommand{\brk@altname}[3]{\expandafter\def\csname#2\expandafter\@gobble\string#1\endcsname{#1[#3]}}
\newcommand{\brk@usearg}[3]{%
  \def\brk@star{*}\def\brk@blank{}\def\brk@arg{#1}%
  \ifx\brk@arg\brk@blank\def\brk@arg{brk@ord}\fi%
  \ifx\brk@arg\brk@star\def\brk@arg{brk@var}\fi%
  \csname\brk@arg #2\endcsname#3}

\newcommand{\DeclareMathBrackets}[3]{
  \newcommand{#1}[2][]{\brk@usearg{##1}{l}{#2}##2\brk@usearg{##1}{r}{#3}}
  \brk@altname{#1}{big}{big}\brk@altname{#1}{lr}{*}}
\newcommand{\DeclareMathBiBrackets}[4]{
  \newcommand{#1}[3][]{\brk@usearg{##1}{l}{#2}##2#3##3\brk@usearg{##1}{r}{#4}}
  \brk@altname{#1}{big}{big}\brk@altname{#1}{lr}{*}}
\newcommand{\DeclareMathBiMBracketsStar}[4]{
  \newcommand{#1}[3][]{\brk@usearg{##1}{l}{#2}##2\brk@usearg{##1}{m}{#3}##3\brk@usearg{##1}{r}{#4}}
  \brk@altname{#1}{bi}{big}}
\newcommand{\DeclareMathBiBracketsStar}[4]{
  \newcommand{#1}[3][]{\brk@usearg{##1}{l}{#2}##2\brk@usearg{##1}{}{#3}##3\brk@usearg{##1}{r}{#4}}
  \brk@altname{#1}{big}{big}}

\makeatother

\DeclareMathBrackets{\brk}{(}{)}
\DeclareMathBrackets{\sbrk}{[}{]}
\DeclareMathBrackets{\set}{\{}{\}}
\DeclareMathBrackets{\abs}{|}{|}
\DeclareMathBrackets{\eval}{.}{|}
\DeclareMathBrackets{\spn}{\langle}{\rangle}
\DeclareMathBiBrackets{\comm}{[}{,}{]}
\DeclareMathBiBrackets{\acomm}{\{}{,}{\}}
\DeclareMathBiBrackets{\gcomm}{[}{,}{\}}

\DeclareMathOperator{\tr}{tr}

\def\[{\begin{equation}}
\def\]{\end{equation}}

\providecommand{\href}[2]{#2}

\makeatletter
\def\mr@ignsp#1 {\ifx\:#1\@empty\else #1\expandafter\mr@ignsp\fi}%
\newcommand{\multiref}[1]{\begingroup%\let\protect\string%
\xdef\mr@no@sparg{\expandafter\mr@ignsp#1 \: }%
\def\mr@comma{}%
\@for\mr@refs:=\mr@no@sparg\do{\mr@comma\def\mr@comma{,}\ref{\mr@refs}}%
\endgroup}
\renewcommand{\eqref}[1]{(\multiref{#1})}
\makeatother

\makeatletter
\newcommand{\namedref}[2]{\hyperref[#2]{#1~\ref*{#2}}}
\newcommand{\secref}{\@ifstar{\namedref{Section}}{\namedref{Sec.}}}
\newcommand{\appref}{\@ifstar{\namedref{Appendix}}{\namedref{App.}}}
\newcommand{\tabref}{\@ifstar{\namedref{Table}}{\namedref{Tab.}}}
\newcommand{\figref}{\@ifstar{\namedref{Figure}}{\namedref{Fig.}}}
\makeatother

\usepackage{mathtools}

\DeclarePairedDelimiter\ket{\lvert}{\rangle}
\DeclarePairedDelimiterX\braket[2]{\langle}{\rangle}{#1 \delimsize\vert #2}

\providecommand{\hypersetup}[1]{}

\hypersetup{plainpages=false}
\hypersetup{pdfpagemode=UseNone}
\hypersetup{bookmarksnumbered=true}
\hypersetup{pdfstartview=FitH}
\hypersetup{colorlinks=false}
\hypersetup{citebordercolor={.5 1 .5}}
\hypersetup{urlbordercolor={.5 1 1}}
\hypersetup{linkbordercolor={1 .7 .7}}
\makeatletter
\let\@keywords\@empty
\let\@subject\@empty
\providecommand{\keywords}[1]{\gdef\@keywords{#1}}
\providecommand{\subject}[1]{\gdef\@subject{#1}}
\def\thetitle{\@title}
\def\theauthor{\@author}
\def\thesubject{\@subject}
\def\thedate{\@date}
\def\thekeywords{\@keywords}
\makeatother
\AtBeginDocument{
\hypersetup{pdftitle={\thetitle}}%
\hypersetup{pdfauthor={\theauthor}}%
\hypersetup{pdfsubject={\thesubject}}%
\hypersetup{pdfkeywords={\thekeywords}}%
}

\title{New integrable 1D models of superconductivity}
\author{ Marius de Leeuw, Anton Pribytok, Ana L. Retore and Paul Ryan}

\begin{document}

\pdfbookmark[1]{Title Page}{title}
\thispagestyle{empty}

%\begingroup\raggedleft\footnotesize\ttfamily
%\arxivlink{yymm.nnnn}
%\par\endgroup

\vspace*{2cm}
\begin{center}%
\begingroup\Large\bfseries\thetitle\par\endgroup
\vspace{1cm}

\begingroup\scshape\theauthor\par\endgroup
\vspace{5mm}%

\begingroup\itshape
School of Mathematics
\& Hamilton Mathematics Institute\\
Trinity College Dublin\\
Dublin, Ireland
\par\endgroup
\vspace{5mm}

\begingroup\ttfamily
$\{$mdeleeuw,
apribytok,
retorea,
pryan$\}$@maths.tcd.ie
\par\endgroup

\vfill

\textbf{Abstract}\vspace{5mm}

\begin{minipage}{12.7cm}
In this paper we find new integrable one-dimensional lattice models of electrons. We describe all such nearest-neighbour integrable models with $\alg{su}(2)\times\alg{su}(2)$  symmetry by classifying solutions of the Yang-Baxter equation following the procedure first introduced in \cite{deLeeuw:2019zsi}. We find $ 12 $ R-matrices of difference form, some of which can be related to known models such as the XXX spin chain and the free Hubbard model, and some are new models. In addition, integrable generalizations of the Hubbard model are found by keeping the kinetic term of the Hamiltonian and adding all terms which preserve fermion number. We find that most of the new models cannot be diagonalized using the standard nested Bethe Ansatz.
\end{minipage}

\vspace*{4cm}

\end{center}

\newpage

%%%%%%%%%%%%%%%%%%%%%%%%%%%%%%%%%%%%%%%%%%%%%%%%%%%%%%%%%%%%%%%%%%%%%%%%%%%%%%%%
%%%%%%%%%%%%%%%%%%%%%%%%%%%%%%%%%%%%%%%%%%%%%%%%%%%%%%%%%%%%%%%%%%%%%%%%%%%%%%%%

\section{Introduction}

It is important to study strongly correlated electrons to understand physical phenomena such as superconductivity. The prototypical example of a model in which this is possible is the Hubbard model \cite{hubbard1963electron,HubBook} which is a basic model of electrons in the conduction band of a solid. To each site of the solid, we associate a four-dimensional Hilbert space. The site can be either vacant, occupied by a single electron with spin up or down, or by a pair of electrons. The Hubbard model Hamiltonian, $\bH^{(Hub)}$, written in terms of oscillators, is then given by 
\begin{align}
\bH^{(Hub)} =  \sum_i  \sum_{\alpha = \uparrow,\downarrow} (\cdo_{\alpha,i}\co_{\alpha,i+1} + \cdo_{\alpha,i+1}\co_{\alpha,i} )+ \mathfrak{u} \, \no_{\uparrow,i}\no_{\downarrow,i}.\end{align}
The kinetic part describes a hopping term that allows electrons to move to neighboring sites whereas the potential term measures the number of electron pairs on each site and $\mathfrak{u}$ sets the overall scale. 

In the one-dimensional case, it was found that the Hubbard model is integrable \cite{Lieb:1968zza,SriramShastry1988} which means that there is an underlying $R$-matrix, i.e. a solution of the Yang-Baxter equation 
\begin{align}
R_{12}(u,v)R_{13}(u,w)R_{23}(v,w) = R_{23}(v,w)R_{13}(u,w)R_{12}(u,v)
\end{align}
which generates an infinite family of conserved charges which commute with the Hubbard Hamiltonian. Furthermore, the $R$-matrix satisfies the \textit{regularity} condition, $R_{12}(u,u)=P_{12}$, where $P_{12}$ is the permutation operator. 

It is an interesting question whether there are other integrable models that describe similar physical systems as the Hubbard model. Recently a new approach \cite{deLeeuw:2019zsi} has been put forward to classify solutions of the Yang-Baxter equation of difference form meaning the $R$-matrix satisfies $R(u,v)=R(u-v)$. 
The central idea behind this method is to take the Hamiltonian, rather than the $R$-matrix as a starting point. More precisely, by using the so-called boost symmetry to generate the corresponding tower of conserved charges, a correspondence is found between integrable systems and a set of polynomial equations. Once the conserved charges are found, we reconstruct the corresponding $R$-matrix. In this sense, our method is similar to \cite{1994JPhy1...4.1151I} where the Yang-Baxter equation was perturbatively solved for 19-vertex models. 

The main focus of our approach is to classify solutions of the Yang-Baxter equation by using conserved charges. This differs from other approaches focused on the Hamiltonian such as \cite{Frappat1,Frappat2,Frappat3}. In these papers Hamiltonians which are solvable by means of the coordinate Bethe Ansatz are classified. There are various examples of integrable Hamiltonians which are not solvable by coordinate Bethe Ansatz means and indeed for the new models we find in this paper the standard Bethe Ansatz approach does not seem to apply. 

In this paper, we apply the method of \cite{deLeeuw:2019zsi}  to the set of integrable models whose physical space is the aforementioned conduction band and find the corresponding new regular solutions $R(u)$ of the Yang-Baxter equation
\begin{align}\label{YBE}
R_{12}(u-v)R_{13}(u)R_{23}(v) = R_{23}(v)R_{13}(u)R_{12}(u-v).
\end{align}
The full set of such models is very large. A priori, the Hamiltonian has 256 free parameters and solving coupled polynomial systems of equations is a challenging task. 
However in the present setting, the problem becomes more tractable if we impose some further restrictions on our Hamiltonian. The set of models we will consider share some features with the Hubbard model and have a reduced set of free parameters and we will consider two classes of such models. We will first consider models which have  $\alg{su}(2)\times \alg{su}(2)$\footnote{More correctly we are considering the algebra $\alg{su}(2)\oplus\alg{su}(2)$, but have decided to stick to the notation $\alg{su}(2)\times\alg{su}(2)$ which is prevalent in most of the physics literature. } symmetry and then models whose kinetic part is given by the kinetic part of the Hubbard model. However, the Hubbard model itself will not appear as one of our solutions as its $R$-matrix is in fact not of difference form but nevertheless in this way we can construct new integrable models that share many properties with it.

In the Hubbard model the $\alg{su}(2)\times \alg{su}(2)$ algebra is realized as a charge symmetry $\alg{su}_{\mathcal{C}}(2)$ and a spin symmetry $\alg{su}_{\eta}(2)$ \cite{article,Essler:1991wg}. This symmetry can actually be extended to an algebra called centrally extended $\alg{\su}(2|2)$ see \cite{Beisert:2006qh,deLeeuw:2015ula}, which is the symmetry algebra which plays a crucial role in the AdS/CFT correspondence \cite{Beisert:2005tm}.

For this class of models we recover the familiar spin chains whose underlying symmetry algebra contains $\alg{su}(2)\times \alg{su}(2)$, such as the $\alg{su}(4),\ \alg{su}(2|2),\ \alg{sp}(4)$ and the $\alg{so}(4)$ spin chains. In the $\alg{so}(4)$ case, however, we find that the Hamiltonian admits an extra parameter $C$ (see \eqref{eq:Hso4}) which is not present for the usual $\alg{so}(n)$ spin chains. We have checked that the spectrum depends non-trivially on this parameter and it arises from the decomposition of $\alg{so}(4) = \alg{su}(2)\times \alg{su}(2)$.

Apart from these well-known spin chains, we find several new models that seem to have interesting physical properties. In particular we find three models in which only electron pairs can propagate. The fermionic degrees of freedom seem to freeze out, but they affect the spectrum non-trivially. The standard Bethe Ansatz approach breaks down for these models and we have not been able to find an alternative way to compute the spectrum, however promising approaches are proposed in the concluding discussion. We performed a study of the spectrum for small spin chain lengths and low excitation numbers and found a very non-trivial structure. 

For our second class of models, we consider deformations of the free Hubbard model. We keep the kinetic part of the Hubbard model Hamiltonian and we add an arbitrary potential and a possible new hopping term for electron pairs. We allow for the most general deformation which preserves electron number so that we still have a physical interpretation of our model. It turns out that we find four integrable models. Three of those are simple combinations of lower dimensional integrable spin chains in which the electrons with spin up and down decouple. However, we find one new model which has two free parameters which has a very non-trivial Hamiltonian. In particular, it contains a term which flips the spins of electrons, mixing $|\!\uparrow\uparrow\rangle$ with $|\!\downarrow\downarrow\rangle$ just as in the XYZ spin chain. As a consequence, this new model has potentially very interesting physics. It is integrable, but due to the fact that it contains some XYZ type-terms, the standard coordinate Bethe Ansatz cannot be applied. 

This paper is organized as follows. In the first section we will recapitulate the method from \cite{deLeeuw:2019zsi} that we will use and discuss our conventions. In the next two sections we discuss $\alg{su}(2)\times \alg{su}(2)$ symmetric models. After this we give the classification of the second class of integrable models. We end with a discussion and conclusions.

%Many solutions of the Yang-Baxter equation are already known \cite{Jimbo1986} \cite{Bazhanov1987} \cite{Kuniba:1991yd} \cite{Stolin1997} and several new ones were found recently \cite{Yu_2019} \cite{Vieira}\cite{deLeeuw:2019zsi} \cite{Vieira:2019vog}.

%%%%%%%%%%%%%%%%%%%%%%%%%%%%%%%%%%%%%%%%%%%%%%%%%%%%%%%%%%%%%%%%%%%%%%%%%%%%%%%%
%%%%%%%%%%%%%%%%%%%%%%%%%%%%%%%%%%%%%%%%%%%%%%%%%%%%%%%%%%%%%%%%%%%%%%%%%%%%%%%%

\section{Set-up and method}

We employ the method from \cite{deLeeuw:2019zsi} to classify one-dimensional integrable models of electrons. We will consider the set-up similar to that of the Hubbard model, which means that the local Hilbert space is four-dimensional. Each lattice site can be empty, occupied by one electron with spin up or down, or by a pair of electrons. This means we will find $16\times 16$ solutions of the Yang-Baxter equation. A full classification of such models is currently not feasible, but if we impose some symmetry conditions on our Hamiltonian, new solutions of the Yang-Baxter equation can be found.

\subsection{Hamiltonian}

The main idea of  \cite{deLeeuw:2019zsi}  is to consider a general Hamiltonian $\bH$, also denoted $\bQ_2$,
\begin{align}
\bH =\bQ_2= \sum_n \lH_{n,n+1},
\end{align}
where the Hamiltonian density $\lH$ is a $16\times16$ matrix. Then we use the so-called boost operator \cite{10.1143/PTP.70.1508,Tetelman,Loebbert:2016cdm}
\begin{equation}\label{eq:boost}
\lB[\bQ_2]:=\sum_{n=-\infty}^\infty n \lH_{n,n+1}
\end{equation}
to generate the higher conserved charges $\bQ_i$ that are present in integrable systems.
More precisely, the boost operator \cite{10.1143/PTP.70.1508,Tetelman,Loebbert:2016cdm} can be used to recursively generate all conserved charges $\bQ_r$ in the following way 
\begin{equation}
\bQ_{r+1} \sim  [\lB[\bQ_2],\bQ_r].
\end{equation}
By imposing that $0=[\bQ_2,\bQ_3] = [\bQ_3,\bQ_4]=\ldots$ we derive a set of coupled polynomial equations on the coefficients of $\lH$, which we then solve. For the models we consider in this paper it turns out that imposing $[\bQ_2,\bQ_3]=0$ is a sufficient condition. Indeed, for the Hamiltonians corresponding to solutions of $[\bQ_2,\bQ_3]=0$, we can subsequently solve the Yang-Baxter equation. %More precisely, we assume that we can expand the $R$-matrix as
%
%\begin{align}
%R = P + P\lH u+ P\lH^2 \frac{u^2}{2}+ \sum_{n\geq 3}R^{(n)} u^n.
%\end{align}
%
For each of the Hamiltonians that we find, we are consequently able to find a corresponding $R$-matrix, which proves the integrability of the underlying model.

\subsection{Identifications}\label{sec:ident}

As outlined in \cite{deLeeuw:2019zsi}, finding solutions to the Yang--Baxter equation in this way leads to a large redundancy in solutions. In particular, some solutions can be related to each other by simple transformations and we will identify solutions which can be related in this way. The transformations under which we identify solutions are:

\paragraph{Normalization} We can clearly multiply any solution of the Yang-Baxter equation by a scalar function. 

\paragraph{Reparametrisation}
The $R$-matrix will depend on a number of free parameters. In particular, one is free to choose reparametrisations, thus some solutions that we find can be related by a redefinition of the parameters and clearly do not define a different integrable model.

\paragraph{Basis transformation}
Any local basis transformation $V:\mathbb{C}^4\rightarrow\mathbb{C}^4$ can be applied to the $R$-matrix
\begin{align}
R \mapsto R^V = (V\otimes V) R (V^{-1}\otimes V^{-1})
\end{align}
to define a different $R$-matrix which satisfies the Yang-Baxter equation.

\paragraph{Discrete transformations}
It is straightforward to check that if $R(u)$ is a solution of the Yang-Baxter equation then $PR(u)P$ and $R(u)^T$ are solutions as well. This means that transposition and permutation are further discrete transformations that map an integrable Hamiltonian to a different integrable Hamiltonian. 

\paragraph{Twists}
If $[R,V\otimes V]=[R,W\otimes W]=0$ then we can define a twisted model
\begin{align}\label{eq:twist}
R \mapsto R^{V,W} = (V\otimes  W) R (W^{-1} \otimes V^{-1}).
\end{align}
Notice that a twist can affect the symmetry properties of the $R$-matrix since $V$ or $W$ need not commute with the symmetry generators. 

In particular, twists generically alter the physical properties of the integrable model. As such, while we identify models that can be related by twists since their mathematical structures are equivalent, they correspond to different physical models. In this way, any new solutions that we find actually give rise to a wide variety of new, physical integrable models. 

\subsection{Oscillators and graded models} 
Finally, let us recall that, as was remarked in the introduction, the Hubbard model Hamiltonian is often formulated by introducing two sets of fermionic oscillators $\co_{\alpha,j},\cdo_{\alpha,j}$ for each lattice site $j$ with $\alpha =\, \uparrow,\downarrow$ satisfying the usual anti-commutation relations
\begin{align}\label{eq:oscdef}
&\{\cdo_{\alpha,i},\co_{\beta,j} \} = \delta_{\alpha\beta}\delta_{ij}
\end{align}
In terms of these oscillators the Hubbard model Hamiltonian is given by
\begin{align}
\bH^{(Hub)} =  \sum_i  \sum_{\alpha = \uparrow,\downarrow} (\cdo_{\alpha,i}\co_{\alpha,i+1} + \cdo_{\alpha,i+1}\co_{\alpha,i} )+ \mathfrak{u} \, \no_{\uparrow,i}\no_{\downarrow,i}.\end{align}
and we have also introduced the number operators $\no_{\alpha,j} = \cdo_{\alpha,j} \co_{\alpha,j}$. Each $4$-dimensional local Hilbert space $V_j$ is then spanned by 
\begin{align}
&|\phi_1\rangle= \vac,
&&|\phi_2\rangle= \cdo_{\uparrow,j} \cdo_{\downarrow,j} |0\rangle,
&&|\psi_1\rangle = \cdo_{\uparrow,j} |0\rangle,
&&|\psi_2\rangle = \cdo_{\downarrow,j} |0\rangle.
\label{phi12psi12}
\end{align}
where $\ket{0}$ is the vacuum state satisfying $\co_{\alpha,j}\ket{0}=0$. Since our goal in this paper is to construct integrable Hubbard-type models it is natural to also use the oscillator formalism when implementing the $\mathfrak{su}(2)\times \mathfrak{su}(2)$ symmetry we consider, complimenting the traditional matrix approach. 

The oscillators naturally introduce the structure of a graded vector space on each local space. Indeed, it is natural to identify $|\phi_1\rangle$ and $|\phi_2\rangle$ above as even basis vectors and $|\psi_1\rangle$ and $|\psi_2\rangle$ as odd. Hence, each local Hilbert space acquires the structure of the graded vector space $\mathbb{C}^{2|2}$. The integrable structures underlying the Hubbard model, in particular the Yang-Baxter equation, can be naturally modified to account for this graded structure, see for instance \cite{Kulish:1985bj,Bracken:1994hz, batchelor2008quantum,HubBook}. In Appendix \ref{app:Grading} we briefly review this formalism. 

%%%%%%%%%%%%%%%%%%%%%%%%%%%%%%%%%%%%%%%%%%%%%%%%%%%%%%%%%%%%%%%%%%%%%%%%%%%%%%%%
%%%%%%%%%%%%%%%%%%%%%%%%%%%%%%%%%%%%%%%%%%%%%%%%%%%%%%%%%%%%%%%%%%%%%%%%%%%%%%%%

\section{Hubbard type models}

The most general nearest-neightbour Hamiltonian where the local Hilbert space is four-dimensional has 256 components. Fully classifying all integrable solutions is currently not feasible, but we can restrict to a proper subset of physically interesting Hamiltonians with a smaller amount of free parameters. We would like to restrict to models which exhibit  spin and charge $\alg{su}(2)$ symmetry, similar to the Hubbard model \cite{HubBook}. It turns out that there are two non-trivial four-dimensional representations of $\alg{su}(2)\times\alg{su}(2)$, see Appendix \ref{app:su2}. In this section we will consider the case in which the representation can be written as a direct sum which is the case of the Hubbard model. 

\subsection{$\su(2)\times \su(2)$ symmetry}

We consider the four-dimensional representation $\rho_{2\oplus1\oplus1}$ of $\alg{su}(2)\times\alg{su}(2)$ in which both $\alg{su}(2)$'s are represented two-dimensionally, see Appendix \ref{app:su2}.
\begin{align}
\rho_{2\oplus1\oplus1} (t^L_i\times t^R_j) =
\begin{pmatrix}
\rho_2(t^L_i) & 0\\
0 & \rho_2(t^R_i)
\end{pmatrix}.
\end{align}
For any $\mcA\in \su(2)\times \su(2)$, we then demand that 
\begin{align}
[\lH_{12},\rho_{2\oplus1\oplus1}(\mcA)\otimes 1+1\otimes \rho_{2\oplus1\oplus1}(\mcA)]=0.
\end{align}
Examples of models that have this symmetry are the $\mathrm{AdS}_5\times\mathrm{S}^5$ superstring, the Hubbard model and the $\alg{su}(4)$ Heisenberg spin chain. However, only the last model has an $R$-matrix which is of difference form.

\paragraph{Hamiltonian} It is straightforward to show that an $\su(2)\times \su(2)$ invariant Hamiltonian in the above sense takes the form
\begin{align}
\lH |\phi_a \phi_b \rangle &= A  |\phi_a \phi_b \rangle + B  |\phi_b \phi_a \rangle + C \epsilon_{ab}\epsilon^{\alpha\beta} |\psi_\alpha \psi_\beta \rangle, \label{Hphiphi}\\
\lH |\phi_a \psi_\beta \rangle &= G  |\phi_a \psi_\beta \rangle + H  |\psi_\beta \phi_a \rangle, \label{Hphipsi} \\
\lH |\psi_\alpha \phi_b \rangle &= K  |\psi_\alpha \phi_b \rangle + L  |\phi_b \psi_\alpha \rangle,  \label{Hpsiphi}\\
\lH |\psi_\alpha \psi_\beta \rangle &= D  |\psi_\alpha \psi_\beta \rangle + E  |\psi_\beta \psi_\alpha \rangle + F \epsilon^{ab}\epsilon_{\alpha\beta} |\phi_a \phi_b \rangle. \label{Hpsipsi}
\end{align}
Here $\phi_{1,2}$ and  $\psi_{1,2}$ span the two independent $\alg{su}(2)$ fundamental representations. Explicitly in matrix form, the Hamiltonian density is given by
\setcounter{MaxMatrixCols}{20}
\begin{align}\label{eq:Hsu2}
\lH = \tiny{
\begin{pmatrix}
A+B & 0 & 0 & 0 & 0 & 0 & 0 & 0 & 0 & 0 & 0 & 0 & 0 & 0 & 0 & 0 \\
 0 & A & 0 & 0 & B & 0 & 0 & 0 & 0 & 0 & 0 & F & 0 & 0 & -F & 0 \\
 0 & 0 & G & 0 & 0 & 0 & 0 & 0 & L & 0 & 0 & 0 & 0 & 0 & 0 & 0 \\
 0 & 0 & 0 & G & 0 & 0 & 0 & 0 & 0 & 0 & 0 & 0 & L & 0 & 0 & 0 \\
 0 & B & 0 & 0 & A & 0 & 0 & 0 & 0 & 0 & 0 & -F & 0 & 0 & F & 0 \\
 0 & 0 & 0 & 0 & 0 & A+B & 0 & 0 & 0 & 0 & 0 & 0 & 0 & 0 & 0 & 0 \\
 0 & 0 & 0 & 0 & 0 & 0 & G & 0 & 0 & L & 0 & 0 & 0 & 0 & 0 & 0 \\
 0 & 0 & 0 & 0 & 0 & 0 & 0 & G & 0 & 0 & 0 & 0 & 0 & L & 0 & 0 \\
 0 & 0 & H & 0 & 0 & 0 & 0 & 0 & K & 0 & 0 & 0 & 0 & 0 & 0 & 0 \\
 0 & 0 & 0 & 0 & 0 & 0 & H & 0 & 0 & K & 0 & 0 & 0 & 0 & 0 & 0 \\
 0 & 0 & 0 & 0 & 0 & 0 & 0 & 0 & 0 & 0 & D+E & 0 & 0 & 0 & 0 & 0 \\
 0 & C & 0 & 0 & -C & 0 & 0 & 0 & 0 & 0 & 0 & D & 0 & 0 & E & 0 \\
 0 & 0 & 0 & H & 0 & 0 & 0 & 0 & 0 & 0 & 0 & 0 & K & 0 & 0 & 0 \\
 0 & 0 & 0 & 0 & 0 & 0 & 0 & H & 0 & 0 & 0 & 0 & 0 & K & 0 & 0 \\
 0 & -C & 0 & 0 & C & 0 & 0 & 0 & 0 & 0 & 0 & E & 0 & 0 & D & 0 \\
 0 & 0 & 0 & 0 & 0 & 0 & 0 & 0 & 0 & 0 & 0 & 0 & 0 & 0 & 0 & D+E
\end{pmatrix}}.
\end{align}

\paragraph{Oscillator representation}

We can define our $\su(2)\times \alg{su}(2)$ representation $\rho_{osc}$ in the oscillator language 
\begin{align}
&\rho_{osc} (t^L_1)  = \frac{1}{2}\left(\cdo_\uparrow \cdo_\downarrow  + \co_\uparrow \co_\downarrow\right)
&&\rho_{osc} (t^L_2)  =\frac{i}{2}\left( \cdo_\uparrow \cdo_\downarrow  - \co_\uparrow \co_\downarrow\right)
&&\rho_{osc} (t^L_3)   = \frac{i}{2}\left(\no_\uparrow + \no_\downarrow -1\right), \\
&\rho_{osc} (t^R_1)  = \frac{1}{2}\left(\cdo_\uparrow \co_\downarrow  + \co_\uparrow \cdo_\downarrow\right)
&&\rho_{osc} (t^R_2)  = \frac{i}{2}\left(\cdo_\uparrow \co_\downarrow - \co_\uparrow \cdo_\downarrow\right)
&&\rho_{osc} (t^R_3)   = -\frac{i}{2}\left(\no_\downarrow - \no_\uparrow .\right) 
%&\mathbb{E}_1 =\cdo_\uparrow \cdo_\downarrow,
%&&\mathbb{F}_1 = \co_\uparrow \co_\downarrow,
%&&\mathbb{H}_1 = \no_\uparrow + \no_\downarrow -1, \\
%&\mathbb{E}_2 =\cdo_\uparrow \co_\downarrow,
%&&\mathbb{F}_2 =\co_\uparrow \cdo_\downarrow,
%&&\mathbb{H}_2 = \no_\downarrow - \no_\uparrow .
\end{align}
It is straightforward to check from the defining anti-commutation relations of the oscillators \eqref{eq:oscdef} that these operators satisfy the $\su(2)$ defining relations.

The most general two-site operator which commutes with both of the above $\su(2)$ oscillator representations again has 10 free parameters $C_i$ and is given by
\begin{align}
\lH_{12} =
&~\sum_{\alpha\neq\beta}\Big[ ( \cdo_{\alpha,1}\co_{\alpha,2} + \co_{\alpha,1} \cdo_{\alpha,2})(C_1 + C_2 (\no_{\beta,1} - \no_{\beta,2})^2 ) + \nonumber\\
&\qquad ( \cdo_{\alpha,1}\co_{\alpha,2} - \co_{\alpha,1} \cdo_{\alpha,2})(C_3(\no_{\beta,1}-\half) + C_4 (\no_{\beta,2}-\half) )\Big]\nonumber \\
&~+( \cdo_{\uparrow,1}\cdo_{\downarrow,1}\co_{\uparrow,2}\co _{\downarrow,2} +  \co_{\uparrow,1}\co_{\downarrow,1}\cdo_{\uparrow,2}\cdo _{\downarrow,2}) C_5 + \nonumber
( \cdo_{\uparrow,1}\co_{\downarrow,1}\cdo_{\downarrow,2}\co _{\uparrow,2} +  \cdo_{\downarrow,1}\co_{\uparrow,1}\cdo_{\uparrow,2}\co _{\downarrow,2}) C_6 \nonumber\\
&~ + C_7 (\no_{\uparrow,1}-\half)(\no_{\downarrow,1}-\half) + C_8 (\no_{\uparrow,2}-\half)(\no_{\downarrow,2}-\half)\nonumber  \\
&~ + C_9 (\no_{\uparrow,1}-\no_{\downarrow,1})^2 (\no_{\uparrow,2}-\no_{\downarrow,2})^2 +\nonumber\\
&~ + (C_5-C_6) (\no_{\uparrow,1}\no_{\downarrow,1}+\no_{\uparrow,2}\no_{\downarrow,2}-1)(\no_{\uparrow,1}-\no_{\uparrow,2})(\no_{\downarrow,1}-\no_{\downarrow,2})\nonumber \\ 
&~ +\half C_5 ((\no_{\uparrow,1}-\no_{\downarrow,2})^2+(\no_{\downarrow,1}-\no_{\uparrow,2})^2)+C_0,
\end{align}
\noindent
where 

\begin{align}
& C_0=\frac{1}{2}(B+G+K),\,C_1=\frac{1}{2}(L-H),\,C_2=\frac{1}{2}(C-F+H-L), \nonumber\\
& C_3=\frac{1}{2}(H+L-C-F),\,C_4=\frac{1}{2}(C+F+H+L),\,C_5=-B,\,C_6=E,\nonumber\\
& C_7=2A+B-2K,\, C_8=2A+B-2G,\,C_9=A+B+D+E-G-K.
\end{align}

\subsection{Solutions}

Following the steps in \cite{deLeeuw:2019zsi}, we take a Hamiltonian of the form \eqref{eq:Hsu2} and compute the corresponding density $\lQ_3$ for the next conserved charge. Next, we impose that $[\bQ_2,\bQ_3]=0$ and find a set of coupled cubic polynomial equations. Solving this set of equations leads to 45 solutions, which, after identifying solutions according to the transformations discussed in Section \ref{sec:ident}, results in 12 independent solutions which are listed in Table \ref{tab:table1}.  Each of these models is integrable and we will present the corresponding R-matrices in the next section. Five of these models are new integrable models and we will highlight their properties in the next sections.

\setlength{\tabcolsep}{0.5em} % for the horizontal padding
{\renewcommand{\arraystretch}{1.2}% for the vertical padding
\begin{table}[h]
  \begin{center}
    \begin{tabular}{c||c|c|c|c|c|c|c|c|c|c|} 
      \textbf{ Model} & \textbf{A} & \textbf{B} & \textbf{C}& \textbf{D} & \textbf{E} &\textbf{F}& \textbf{G}& \textbf{H}& \textbf{K}& \textbf{L}\\
      \hline
1 & 0 & 0 & 0 & 0 & 0 & 0 & $a$ & $b$  & $c$ & $d$ \\ \hline
2 &  0 & 0 & 0 & $a+c$ & 0 & 0 & $a$ & $b$ & $c$ & $d$\\ \hline
3 &  0 & 0 & 0 & $a$ & 0 & 0 & $b$ & 0 & $c$ & 0 \\ \hline
4 & $\rho$ &$ -\rho $& 0 & 0 & 0 & 0 & $a $& $\rho e^{-\phi} $& $2 \rho-a$  & $\rho e^\phi$ \\ \hline
5 & $\rho$ &$ -\rho $& 0 & $\rho$ & $-\rho$ & 0 & $a $& $\rho e^{-\phi} $& $2 \rho-a$  & $\rho e^\phi$ \\ \hline
6 & 0 & 0 & 0 & $\rho$ & $\rho$ & 0 & $a$ & $\rho e^{-\phi} $& $2 \rho-a$  & $\rho e^\phi$ \\ \hline
7 & $\rho$ &$ -\rho $& 0 & $\rho$ & $\rho$ & 0 & $a $& $\rho e^{-\phi} $& $2 \rho-a$  & $\rho e^\phi$ \\ \hline
8 & $ \rho$ &$ -\rho $ &$ \rho e^{-\phi} $&$ -\rho$&$ \rho $&$ -\rho e^\phi$ & 0 & 0 & 0 & 0 \\ \hline
9 & $ \rho$ &$ -\rho $ &$ \rho e^{-\phi} $&$ \rho$&$ -\rho $&$ \rho e^\phi$ & 0 & 0 & 0 & 0 \\ \hline
10 & $ \frac{7}{4}\rho$ &$ -\rho $ &$ \half\rho e^{-\phi} $&$ \frac{7}{4}\rho$&$ -\rho $&$ \half\rho e^\phi$ & 0 & 0 & 0 & 0 \\ \hline
11 & $ \rho$ &$ -\rho $ &$ \half\rho e^{-\phi} $&$ \rho$&$ -\rho $&$ \half\rho e^\phi$ & $\sfrac{3}{2}\rho $& $-\sfrac{3}{2}\rho $ & $\sfrac{3}{2}\rho$ & $-\sfrac{3}{2}\rho $ \\ \hline
12 & 0 & 0 & $ - \rho e^{-\phi}$ & 0 & 0 & $\rho  e^{\phi} $ & 0 & $\rho $ & 0 & $ -\rho $\\ \hline
    \end{tabular}
  \end{center}    
\caption{All non-graded integrable spin chains with charge and spin $\alg{su}(2)$ symmetry.}
\label{tab:table1}
\end{table}

\subsection{R-matrices}
As a result of the $\mathfrak{su}(2)\times \mathfrak{su}(2)$ symmetry, all of the $R$-matrices corresponding to the Hamiltonians listed above can be expressed as 
\begin{equation}
{\tiny R_{12}(u)=\left(
\begin{array}{cccccccccccccccc}
 r_1+r_2 & 0 & 0 & 0 & 0 & 0 & 0 & 0 & 0 & 0 & 0 & 0 & 0 & 0 & 0 & 0 \\
 0 & r_1 & 0 & 0 & r_2 & 0 & 0 & 0 & 0 & 0 & 0 & -r_8 & 0 & 0 & r_8 & 0 \\
 0 & 0 & r_4 & 0 & 0 & 0 & 0 & 0 & r_{10} & 0 & 0 & 0 & 0 & 0 & 0 & 0 \\
 0 & 0 & 0 & r_4 & 0 & 0 & 0 & 0 & 0 & 0 & 0 & 0 & r_{10} & 0 & 0 & 0 \\
 0 & r_2 & 0 & 0 & r_1 & 0 & 0 & 0 & 0 & 0 & 0 & r_8 & 0 & 0 & -r_8 & 0 \\
 0 & 0 & 0 & 0 & 0 & r_1+r_2 & 0 & 0 & 0 & 0 & 0 & 0 & 0 & 0 & 0 & 0 \\
 0 & 0 & 0 & 0 & 0 & 0 & r_4 & 0 & 0 & r_{10} & 0 & 0 & 0 & 0 & 0 & 0 \\
 0 & 0 & 0 & 0 & 0 & 0 & 0 & r_4 & 0 & 0 & 0 & 0 & 0 & r_{10} & 0 & 0 \\
 0 & 0 & r_7 & 0 & 0 & 0 & 0 & 0 & r_3 & 0 & 0 & 0 & 0 & 0 & 0 & 0 \\
 0 & 0 & 0 & 0 & 0 & 0 & r_7 & 0 & 0 & r_3 & 0 & 0 & 0 & 0 & 0 & 0 \\
 0 & 0 & 0 & 0 & 0 & 0 & 0 & 0 & 0 & 0 & r_5+r_6 & 0 & 0 & 0 & 0 & 0 \\
 0 & -r_9 & 0 & 0 & r_9 & 0 & 0 & 0 & 0 & 0 & 0 & r_5 & 0 & 0 & r_6 & 0 \\
 0 & 0 & 0 & r_7 & 0 & 0 & 0 & 0 & 0 & 0 & 0 & 0 & r_3 & 0 & 0 & 0 \\
 0 & 0 & 0 & 0 & 0 & 0 & 0 & r_7 & 0 & 0 & 0 & 0 & 0 & r_3 & 0 & 0 \\
 0 & r_9 & 0 & 0 & -r_9 & 0 & 0 & 0 & 0 & 0 & 0 & r_6 & 0 & 0 & r_5 & 0 \\
 0 & 0 & 0 & 0 & 0 & 0 & 0 & 0 & 0 & 0 & 0 & 0 & 0 & 0 & 0 & r_5+r_6 \\
\end{array}
\right)}
\end{equation}
where we have omitted the $u$-dependence on the functions $r_j(u)$ in order to avoid overly bulky expressions. In order to find the $R$-matrix explicitly we must solve the YBE. To this end, we recall that we can express the $R$-matrix in terms of the Hamiltonian density $\lH$ as 
\begin{equation}\label{rperturb}
R(u)=P + P\lH u+P\lH^2\frac{u^2}{2}+\mathcal{O}(u^3).
\end{equation}
The first two orders are fixed by regularity and the definition of the Hamiltonian density. The last order follows by solving the YBE perturbatively, see \textit{e.g.} \cite{1994JPhy1...4.1151I}. Knowing this expansion greatly simplifies solving the YBE. Indeed, already at second order in $u$ it can become apparent that certain entries in $R$ may be equal, or related by a sign change or overall factor. This allows one to consider a reduced ansatz for $R$ where these identifications are introduced. We then attempt to solve the reduced system of functional equations. Specifically, we consider the YBE \eqref{YBE} and differentiate wrt $v$ and evaluate the result at $v=0$. Consistency with \eqref{rperturb} then places initial conditions on the $r_j$ and their derivatives and we can subsequently solve the resulting ODEs.

For each of the 12 models listed above we will simply list the corresponding non-zero functions $r_j(u)$ only. All the R-matrices presented below are regular (i.e. satisfy $ R(0)=P $) and satisfy braiding unitarity $ R_{12}(u)R_{21}(-u)=1 $.
\paragraph{Model 1} For this model, we begin by considering the case where all parameters are generic.
In this case it is convenient to introduce the parameter $\eta$ and the function $g(u)$ defined by 
\begin{equation}
d=\frac{(a+c)^2 \csc^2(\eta )}{4 b},\quad g(u)={\rm arccot}(\tan (\eta ))-\frac{1}{2} u (a+c) \cot (\eta ).
\end{equation}
Then we have
\begin{equation}
\begin{array}{l}
r_2=r_6=1\\
r_3=\frac{a+c}{2b} (\cot (\eta ) \cot (g(u))-1)\\
r_4=\frac{2 b}{a+c} \sin (\eta ) \csc (g(u)) \cos (g(u)+\eta )
\end{array},\quad
\begin{array}{l}
r_7=e^{\frac{1}{2} u (a-c)} \cos (\eta )  \csc (g(u))\\
r_{10}=e^{- u (a-c)}\ r_7
\end{array} 
\end{equation}
Clearly there are two degenerate cases of the above parameters, namely when $b=0$ or $a+c=0$. Hence we must treat these two cases separately. For $b=0$ we have
\begin{equation}
\begin{array}{l}
r_2=r_6=1\\
r_3=\frac{d \left(e^{u (a+c)}-1\right)}{a+c}\\
\end{array},\quad
\begin{array}{l}
r_7=e^{au}\\
r_{10}=e^{cu}
\end{array} 
\end{equation}
whereas for $a+c=0$ we have
\begin{equation}
\begin{array}{l}
r_2=r_6=1\\
r_3=\frac{d}{b}r_4 = \sqrt{\frac{d}{b}} \tan \left(\sqrt{b\ d}\ u\right)\\
\end{array},\quad
\begin{array}{l}
r_7=e^{a u} \sec \left(\sqrt{b\ d}\ u\right)\\
r_{10}=e^{-2au}\ r_7
\end{array} 
\end{equation}

\paragraph{Model 2} 

This model also has the degenerate cases $b=0$ and $a+c=0$, which must be treated separately. For the case when all parameters are generic it is again useful to introduce a parameter $\eta$ and functions $g(u)$ and $h(u)$ defined by
\begin{equation}
\begin{array}{c}
d=\frac{(a+c)^2}{4b} \text{sech}^2(\eta )\\
g(u)=\text{csch}\left(\eta -\frac{1}{2} u (a+c) \tanh (\eta )\right)\\
h(u)=\sinh \left(\frac{1}{2} u (a+c) \tanh (\eta )\right)
\end{array}
\end{equation}
We then have
\begin{equation}
\begin{array}{c}
\begin{split}
& r_2=1\\
& r_3=\frac{2 b}{a+c} g(u)h(u) \text{sech}(\eta )\\
& r_4 =\frac{a+c}{2 b} g(u)h(u) \cosh(\eta)\\
\end{split}
\end{array},\ 
\begin{array}{c}
\begin{split}
& r_6=g(u) \sinh \left(\frac{1}{2} u (a+c) \tanh (\eta )+\eta \right)\\
& r_7=e^{\frac{1}{2} u (a-c)}g(u) \sinh (\eta ) \\
& r_{10} =e^{\frac{1}{2} u (c-a)}g(u) \sinh (\eta ) \\
\end{split}
\end{array}
\end{equation}
When $a+c=0$ we have that $d=0$ and we obtain the same degenerate model as obtained from model 1. When $b=0$ then we obtain
\begin{equation}
\begin{array}{c}
\begin{split}
& r_2=1\\
& r_3=\frac{d}{a+c} \left(e^{u (a+c)}-1\right)\\
& r_6=e^{(a+c)u}\\
\end{split}
\end{array},\ 
\begin{array}{c}
\begin{split}
& r_7=e^{au}\\
& r_{10}=e^{cu}
\end{split}
\end{array}
\end{equation}
\paragraph{Model 3} 
\begin{equation}
\begin{array}{l}
r_2=1\\
r_6=e^{au}\\
\end{array},\quad
\begin{array}{l}
r_7=e^{bu}\\
r_{10}=e^{cu}\\
\end{array} 
\end{equation}
\paragraph{Model 4} 
\begin{equation}
\begin{array}{l}
r_1=-u\rho\ r_2\\
r_2=(1-u\rho)^{-1}\\
r_3=-e^\phi\ r_1\\
r_6=1\\
\end{array},\quad
\begin{array}{l}
r_4=-e^{-\phi}\ r_1\\
r_7=e^{u(a-\rho)}\ r_2\\
r_{10}=e^{u(\rho-a)}\ r_2\\
\end{array}  
\end{equation}
\paragraph{Model 5} 
\begin{equation}
\begin{array}{l}
r_1=r_5=-u\rho\ r_2\\
r_2=r_6=(1-u\rho)^{-1}\\
r_3=-e^\phi\ r_1
\end{array},\quad
\begin{array}{l}
r_4=-e^{-\phi}\ r_1\\
r_7=e^{u(a-\rho)}\ r_2\\
r_{10}=e^{u(\rho-a)}\ r_2\\
\end{array} 
\end{equation}
\paragraph{Model 6} 
\begin{equation}
\begin{array}{l}
r_2=1\\
r_6=(1-u \rho)^{-1}\\
r_5=u \rho\ r_6\\
r_3=e^{\phi}\ r_5
\end{array},\quad
\begin{array}{l}
r_4=e^{-\phi}\ r_5\\
r_7= e^{u(a-\rho)}\ r_6\\
r_{10}=e^{u(\rho-a)}\ r_6
\end{array} 
\end{equation}
\paragraph{Model 7} 
\begin{equation}
\begin{array}{l}
r_1=-r_5=-u\rho\ r_2\\
r_2=r_6=(1-u \rho)^{-1}\\
r_7=e^{u(a-\rho)}r_2
\end{array},\quad
\begin{array}{l}
r_3=-e^{\phi}\ r_1 \\
r_4 = - e^{-\phi}\ r_1\\
r_{10}=e^{u(\rho-a)}r_2
\end{array} 
\end{equation}
\paragraph{Model 8} 
\begin{equation}
\begin{array}{l}
r_1=-r_5=-{\rm tan}(u\ \rho) \\
r_2=1-r_1\\
r_6=1+r_1
\end{array},\quad
\begin{array}{l}
r_7=r_{10}=1\\
r_8 = e^\phi\ r_1 \\
r_9 = - e^{-\phi}\ r_1\\
\end{array} 
\end{equation}
\paragraph{Model 9} 
\begin{equation}
\begin{array}{l}
r_1=r_5 \\
r_2=r_6=1-r_1\\
r_7=r_{10}=1
\end{array},\quad
\begin{array}{l}
r_8 = -e^\phi\ r_1 \\
r_9 = - e^{-\phi}\ r_1
\end{array} 
\end{equation}
$$
r_1 = 2+\sqrt{3} \coth \left(\sqrt{3} \rho  u+\log \left(2-\sqrt{3}\right)\right)
$$
\paragraph{Model 10}
\begin{equation}
\begin{array}{l}
r_1=r_5 = \frac{2(e^{\frac{3 \rho  u}{2}}-1)}{e^{\frac{3 \rho  u}{2}}-4}\\
r_7=r_{10}=e^{-\frac{1}{4} (3 \rho  u)}\\
\end{array},\quad
\begin{array}{l}
r_2=r_6=-\frac{e^{\frac{3 \rho  u}{2}}+2}{e^{\frac{3 \rho  u}{2}}-4}\\
e^{-2\phi}r_9=r_8=-\frac{1}{2}e^{\frac{3 \rho  u}{4}+\phi }\ r_1
\end{array}
\end{equation}
\paragraph{Model 11}
\begin{equation}
\begin{array}{l}
r_1=r_5=\rho  u (3 \rho  u-4)\ f(u) \\
r_2=r_6= 4(1- \rho  u)\ f(u)\\
r_3=r_4=-\frac{3}{2}u\rho\ r_7
\end{array},\quad
\begin{array}{l}
r_7=r_{10}=-2( 3 \rho  u-2)^{-1} \\
e^{2\phi}r_9=r_8= 2 \rho  u e^{\phi }\ f(u)\\
f(u)^{-1}=(\rho  u-2) (3 \rho  u-2)
\end{array} 
\end{equation} 
\paragraph{Model 12} 
\begin{equation}
\begin{array}{l}
r_1=r_5=r_4^2 \\
r_3=-r_4=-{\rm tanh}(u\rho)\\
r_8=e^\phi\ r_4\ r_7
\end{array},\quad
\begin{array}{l}
r_2=r_6=r_7^2 \\
r_7=r_{10}={\rm sech}(u\rho)\\
r_9=-e^{-\phi}\ r_4\ r_7
\end{array} 
\end{equation}

\subsection{Interpretation of models}

In this section we discuss the models that we have listed in Table \ref{tab:table1}. We will relate some of them to known models and discuss some of the properties of the new models. We find two new classes of integrable models. Models 4 and 6 are variations of XXX type models. Models 8,9 and 10 are an exciting new class of models in which only electron pairs can propagate. We will only briefly mention the already known models.

\subsubsection{Parameters}

Most Hamiltonians depend on some parameters corresponding to basis transformations and twists. These parameters are useful when comparing to known models. 
In particular, the Hamiltonians for models 4, 5, 6 and 7 all depend on a parameter $a$ and can be written as follows

\begin{equation}
\lH=\lH_0+\alpha\otimes 1-1\otimes \alpha
\end{equation} 
\noindent
with 
\begin{equation}
\alpha=\text{Diag}(1,1,1-a,1-a)
\end{equation}
\noindent
and $\lH_0 $ is the Hamiltonian for $ a=0 $. Notice that the term $  \alpha\otimes 1-1\otimes \alpha $ does not affect the spectrum for closed spin chains. The $a$ dependence
can be recovered by applying a simple transformation to the $R$-matrix. Define
\begin{equation}
U(u)=\text{Diag}(1,1,e^{au},e^{au}).
\end{equation}
Then 
\begin{equation}
R_{12}(u)=(U(u)\otimes 1) R_{12}(u)|_{a=0}(U(-u)\otimes 1).
\end{equation}
Moreover, the Hamiltonian for models 4, 5, 6 and 7 also all depend on $\phi$, which corresponds to a particular twist
\begin{equation}
\lH(\phi)=\left(G_1\otimes 1\right)\lH(\phi=0)\left(G_1^{-1}\otimes 1\right),
\end{equation}
\noindent
with 
\begin{equation}
G_1=\text{Diag}(1,1,e^{-\phi},e^{-\phi}).
\end{equation}
Keeping this in mind, we can set $a,\phi$ to some convenient values in order to compare with known models in the literature, since the general $a,\phi$ dependence can be easily restored.

Analogously, for models 8, 9, 10, 11 and 12 the parameter $\phi$ corresponds to a rescaling of certain basis elements

\begin{equation}
\lH(\phi)=\left(G_2\otimes G_2\right)\lH(\phi=0)\left(G_2^{-1}\otimes G_2^{-1}\right),
\end{equation}
\noindent
where
\begin{equation}
G_2=\text{Diag}(1,1,e^{-\phi},1).
\end{equation}
Hence $\phi$ will not affect the spectrum and can be accounted for by a simple local basis transformation.

\subsubsection{New models of XXX type}

\paragraph{Model 4} After setting $a=-1,\phi=i\pi$ and $\rho=-1$, we find that Model 4 seems to be a modified version of the $\alg{su}(4)$ spin chain.  Applying a simple basis transformation which sends the basis vectors $E^i \mapsto E^{5-i}$, we can write
\begin{align}\label{eq:Ham4}
\lH^{(4)} = 1 - P + \sum_{i,j,k,l=1,2}  \epsilon_{ij}\epsilon^{kl} E^i_k\otimes E^j_l = 1 - P + \sum_{i,j=1,2} ( E^i_i \otimes E^j_j - E^i_j \otimes E^j_i ) ,
\end{align}
where the sum runs over the $i,j,k,l=1,2$. We denote the standard $4\times 4$ matrix unities by $E^i_k$. The last term can actually be interpreted as the Hamiltonian of the XXX spin chain restricted to the first two basis vectors \eqref{eq:Ham4}, because in general the identity and permutation operators can be expressed as $1 = E^i_i$ and $P = E^i_j \otimes E^j_i $. Thus we can write
\begin{align}
\lH^{(4)} = 1 - P - 1_2 + P_2,
\end{align}
where $1_2$ and $P_2$ are the identity and permutation operator on a two-dimensional subspace generated by $E^{1,2}$.

\paragraph{Model 6} 
Models 4 and 6 are related to each other by a grading type transformation. More precisely, we find for $a=1,\phi=i\pi$ and $\rho=1$
\begin{align}
\lH^{(6)} = 1 - P^f + \sum_{i,j,k,l=1,2}  \epsilon_{ij}\epsilon^{kl} E^i_k\otimes E^j_l,
\end{align}
where $P^f$ is now the graded permutation. In other words, the models are simply related by interchanging the graded and ungraded permutation operator.

\paragraph{Properties}
Both models 4 and 6 have a rational $R$-matrix but also have a non-trivial spectrum. Both models have a $\alg{su}(3)$ subchain generated by the local basis vectors $\{|\phi_1\rangle,|\psi_1\rangle,|\psi_2\rangle\}$. If we consider the closed, length $L$ spin chain corresponding to these integrable models, then it is easy to see that the ferromagnetic vacuum $|0\rangle = |\phi_1\ldots \phi_1\rangle$ is one of the states with zero energy
\begin{align}
\bH |0\rangle= 0.
\end{align}
This is also the lowest energy in the system, but the ground state is clearly degenerate.

The remainder of the spectrum depends on the twist $\phi$, but setting it to $\phi=i\pi$ or $\phi=0$, we find a spectrum with both different eigenvalues and degeneracies than the usual $\alg{su}(4)$ spin chain. In fact, at the moment it is unclear how to perform the Bethe Ansatz for these models.

\subsubsection{New models of electron pairs}

Models 8,9, and 10 all correspond to models in which there is no propagation of individual fermions. This can be seen from the fact that $ G=H=K=L=0 $. These elements of the Hamiltonian exactly correspond to processes where bosons and fermions are permuted. As a consequence, it is unclear how to perform a Bethe Ansatz for these models. The reason is that there is no clear magnon-type picture that underlies the spectrum here. However, individual fermions can move along the chain in the presence of electron pairs. This makes for a very complicated, but interesting system of interacting fermions. This is reflected in a non-trivial spectrum, which we have analysed for small length spin chains in Appendix \ref{sec:reducedH}. Moreover, an Ansatz for the spectrum of this model for up to 2 excitations for any number of sites is presented in section \ref{sec:BA}.

The Hamiltonian for models 9 and 10 are Hermitian. For Model 8, the Hamiltonian is a linear combination of an Hermitian and an anti-Hermitian matrix. Nevertheless, the energy eigenvalues always seems to come in complex conjugate pairs. For Model 9, the ground state has energy 0, which is the eigenvalue of the usual ferromagnetic type vacuum. However, for model 10, the ground state has a non-trivial energy and structure. It would be very interesting to understand the physical properties of these models.

\subsubsection{Known models}

The remaining seven models that we find correspond to well-known models.

\paragraph{Model 1}

Model 1 corresponds to a quadruple embedding of an XXZ-type spin chain with Hamiltonian
\begin{align}
\lH^{(XXZ)} = \begin{pmatrix}
0 & 0 & 0 & 0 \\
0 & a & b & 0 \\
0 & d & c & 0 \\
0 & 0 & 0 & 0 
\end{pmatrix}.
\end{align}
One can show that the spectrum of Model 1 corresponds to the spectrum of $\lH^{(XXZ)}$ where each eigenspace has an extra degeneracy factor of 4. The embedding is simply given by restricting to one vector from each of the $\alg{su}(2)$ doublets $\phi_a,\psi_b$.

\paragraph{Model 2}
Similarly, Model 2 is a staggered-type XXZ model with Hamiltonian
\begin{align}
\lH^{(XXZ^\prime)} = \begin{pmatrix}
0 & 0 & 0 & 0 \\
0 & a & b & 0 \\
0 & d & c & 0 \\
0 & 0 & 0 & a+c 
\end{pmatrix},
\end{align}
again realized on  $\{\phi_a,\psi_b\}$.

\paragraph{Model 3}
The Hamiltonian for Model 3 is diagonal and hence it is trivially integrable.

\paragraph{Model 5} Model 5 corresponds to the twisted $\alg{su}(4)$ spin chain. More specifically, if we set $a=-1,\phi=i\pi$ and $\rho=-1$ we recover
\begin{align}
\lH^{(4)} \mapsto 1 - P,
\end{align}
which indeed is the $\alg{su}(4)$ spin chain Hamiltonian.

\paragraph{Model 7} Model 7 corresponds to the twisted $\alg{su}(2|2)$ spin chain \cite{Essler:1992py}. We find that we recover
\begin{align}
\lH^{(4)} \mapsto 1 - P^f,
\end{align}
upon setting $a=1, \phi =i\pi $ and $\rho=1$.

\paragraph{Model 11} Model 11 corresponds to the $ \alg{sp}(4) $ spin chain, \cite{BERG1978125,KAROWSKI1979244,Kulish:1979cr,Reshetikhin:1986vd}
\begin{align}
R^{   \alg{sp}(4)} = u 1 + P - \frac{u}{u+3}(-1)^{p(i) + p(k)} E^{i}_j\otimes E^{5-i}_{5-j}.
\end{align}
It can be shown that 
\begin{align}
&\lH^{(11)} = \frac{3\rho}{2} 1 - (U\otimes U)\lH^{(\alg{sp}(4))} (U^{-1}\otimes U^{-1}), 
&&
U = \begin{pmatrix}
 1 & 0 & 0 & 0 \\
 0 & 0 & 0 & e^{\phi /2} \\
 0 & 0 & 1 & 0 \\
 0 & e^{-\phi /2} & 0 & 0 
\end{pmatrix}.
\end{align}
The Bethe Ansatz for this known model has been worked out for example in \cite{Reshetikhin:1986vd,Li:2018xrb}.

\paragraph{Model 12} Model 12 corresponds to the free Hubbard model, \textit{i.e.} just the kinetic term. In order to see this, we need to consider a twist \eqref{eq:twist} with $V = \mathrm{diag}(1,-1,i,i)$ and $W=1$.
On the level of the Hamiltonian we find
\begin{align}
\lH^{(V)}_{12} =-i  V_1 \lH V_1^{-1}.
\label{twist}
\end{align}
This relation is needed to make contact with the regular Hubbard model, because the charge $\su_{\mathcal{C}}(2)$ is twisted. Moreover, we need to make the model graded. We also find that we can put $\phi=0$ by using a basis transformation, so that
\begin{equation}
\lH^{(Hub) }  \sim  (U\otimes U)D^r (V_1\lH^{(12),f}V_1^{-1})D^r\left(U^{-1}\otimes U^{-1}\right),
\end{equation}
\noindent
with 
\begin{equation}
U=\begin{pmatrix}
0 & e^{-\frac{\phi}{2}} & 0 & 0\\
e^{-\frac{\phi}{2}} & 0 & 0 & 0\\
0 & 0 & i & 0\\
0 & 0 & 0 & i
\end{pmatrix}.
\end{equation}
By this map we see that this model splits into two disjoint XX type spin chains.

%%%%%%%%%%%%%%%%%%%%%%%%%%%%%%%%%%%%%%%%%%%%%%%%%%%%%%%%%%%%%%%%%%%%%%%%%%%%%%%%
%%%%%%%%%%%%%%%%%%%%%%%%%%%%%%%%%%%%%%%%%%%%%%%%%%%%%%%%%%%%%%%%%%%%%%%%%%%%%%%%

\section{Other $\alg{su}(2)\times\alg{su}(2)$ invariant models}

We notice that the $\alg{so}(4) \sim \alg{su}(2)\times\alg{su}(2)$ spin chain is not in our list of spin and charge $\alg{su}(2)$ invariant models listed in Table \ref{tab:table1} . This is due to the fact that there is one further non-trivial four-dimensional representation, $\rho_{2\oplus2}$, of $\alg{su}(2)\times\alg{su}(2)$ which gives rise to the $\alg{so}(4)$ spin chain, see Appendix \ref{app:su2xsu2}. This representation is given by
\begin{align}
&\rho_{2\oplus2} (t^L_i)  = 1 \otimes  \rho_2(t_i) ,
&&\rho_{2\oplus2} (t^R_i)  =  \rho_2(t_i) \otimes 1,
\end{align}
where $t^L \times t^R \in \alg{su}(2)\times\alg{su}(2)$.
It is straightforward to check that the invariant Hamiltonian under the $\rho_{2\oplus2}\sim \rho_{\alg{so}(4)}$ representation takes the general form
\begin{align}
\lH = A + B P + C K + D \epsilon_{ijkl} E^i_k\otimes E^j_l,
\end{align}
where $P$ is the permutation operator and $K = E^i_j \otimes E^i_j$  is the so-called trace operator. We sum over repeated indices. The matrices $E^i_j$ are the standard $4\times4$ matrix unities.

Following the steps from \cite{deLeeuw:2019zsi}, we only find two new integrable Hamiltonians
\begin{align}\label{eq:Hso4}
\lH^{(13)} &= A - B P + B K + C \epsilon_{ijkl} E^i_k\otimes E^j_l,\\
\lH^{(14)} &= A K,
\end{align}
and the corresponding R-matrices are given by
\begin{align}
R^{(13)} &=\frac{(1+ Au)u}{1-Bu}\Big[(u \left(B^2-C^2\right)-B)1 + \frac{1-Bu}{u}P +B K + C \epsilon_{ijkl} E^i_k\otimes E^j_l \Big], \\
R^{(14)} &=(1+ Au)\Big[ \frac{\sqrt{3} \coth \left(\sqrt{3} B u\right)-2}{\sqrt{3} \coth \left(\sqrt{3}
   B u\right)-1}P +\frac{1}{\sqrt{3} \coth \left(\sqrt{3} B u\right)-1} K  \Big] .
\end{align}
For $C=0$, Model 13 corresponds to the usual $\alg{so}(n)$ spin chain for $n=4$ \cite{Tu_2008}. The presence of $C$ corresponds to the fact that exactly in the four-dimensional case there is an extra invariant contraction, where all indices are contracted with the Levi-Civita symbol. The spectrum depends non-trivially on $C$ and it appears due to the isomorphism $\alg{so}(4) \sim \alg{su}(2)\times\alg{su}(2)$. Indeed each $\alg{su}(2)$ subalgebra comes with its own quadratic Casimir. For the usual XXX, spin chain, the Hamiltonian can be written as
\begin{align}
\lH^{(XXX)} = \sum_i \sigma^i \otimes \sigma^i,
\end{align}  
where $\sigma^i$ are the Pauli matrices. In this case we see that this decomposition directly generalizes 
\begin{align}\label{eq:Hso4}
\lH^{(13)} &= A + 2 \sum_i \Big[(B+C) \rho_{2\oplus2} (t^L_i) \otimes \rho_{2\oplus2} (t^L_i) + (B-C) \rho_{2\oplus2} (t^R_i) \otimes \rho_{2\oplus2} (t^R_i)   \Big].
\end{align}
In other words, the $\alg{so}(4)$ spin chain can be written as the sum of two independent XXX spin chains and the spectrum is simply the sum of the energies of the XXX spin chains with the relevant coefficients, see also \cite{Kulish:1979cr,Reshetikhin:1986vd}.

%%%%%%%%%%%%%%%%%%%%%%%%%%%%%%%%%%%%%%%%%%%%%%%%%%%%%%%%%%%%%%%%%%%%%%%%%%%%%%%%
%%%%%%%%%%%%%%%%%%%%%%%%%%%%%%%%%%%%%%%%%%%%%%%%%%%%%%%%%%%%%%%%%%%%%%%%%%%%%%%%

\section{Generalized Hubbard models}

We noticed that Model 12 corresponds to the free Hubbard model. We can use this model as a starting point to see if there are any potentials or interaction terms that can be added to this kinetic term while preserving integrability. In this way we would find new integrable Hubbard like deformations. We know that these new models cannot be $\alg{su}(2)\times\alg{su}(2)$ invariant. We would like to only consider models which we can interpret as a model of electrons moving on a one-dimensional lattice or conduction band. To this end, we will only include terms which preserve fermion number. 

Let $\lK_{Hub}$ denote the kinetic term of the Hubbard model, \textit{i.e.}
\begin{align}
\lK_{Hub} = \sum_{\alpha=\uparrow,\downarrow}( \cdo_{\alpha,1}\co_{\alpha,2} +  \cdo_{\alpha,2} \co_{\alpha,1}).
\end{align}
We add other kinetic/hopping terms which act on two electrons simultaneously. We consider a term which describes the hopping of a pair of electrons $\lK_{pair}$ and a term which flips the spins of the electrons on neighboring sites $\lK_{flip}$
\begin{align}
\lK_{pair} &=  A_1 \cdo_{\uparrow,1}\cdo_{\downarrow,1}\co_{\uparrow,2}\co_{\downarrow,2} + A_2 \cdo_{\uparrow,2}\cdo_{\downarrow,2}\co_{\uparrow,1}\co_{\downarrow,1},\\
\lK_{flip} &=  A_3 \cdo_{\uparrow,1}\cdo_{\downarrow,2}\co_{\downarrow,1}\co_{\uparrow,2} + A_4 \cdo_{\downarrow,1}\cdo_{\uparrow,2}\co_{\uparrow,1}\co_{\downarrow,2}
+  A_5 \cdo_{\uparrow,1}\cdo_{\uparrow,2}\co_{\downarrow,1}\co_{\downarrow,2} + A_6 \cdo_{\downarrow,1}\cdo_{\downarrow,2}\co_{\uparrow,1}\co_{\uparrow,2}.
\end{align}
The $\lK_{flip}$ term violates spin conservation as it contains terms which sends $|\!\uparrow\uparrow\rangle \rightarrow |\!\downarrow\downarrow\rangle$ and $|\!\downarrow\downarrow\rangle \rightarrow |\!\uparrow\uparrow\rangle$.
We finally consider the most general potential term written in terms of number operators
\begin{align}
V =&~ B_1 + B_2\, \no_{\uparrow,1} + B_3\, \no_{\downarrow,1}  + B_4\, \no_{\uparrow,1} \no_{\downarrow,1}  + \nonumber\\
&~B_5\,\no_{\uparrow,2}  + B_6\, \no_{\uparrow,1} \no_{\uparrow,2}  + B_7 \,\no_{\downarrow,1} \no_{\uparrow,2}  + B_8\, \no_{\uparrow,1} \no_{\downarrow,1} \no_{\uparrow,2} +\nonumber\\
&~B_9\,\no_{\downarrow,2}  + B_{10} \,\no_{\uparrow,1} \no_{\downarrow,2}   + B_{11}\, \no_{\downarrow,1} \no_{\downarrow,2}   + B_{12}\, \no_{\uparrow,1} \no_{\downarrow,1} \no_{\downarrow,2}   +\nonumber\\
&~B_{13}\,\no_{\uparrow,2} \no_{\downarrow,2}   + B_{14}\, \no_{\uparrow,1} \no_{\uparrow,2} \no_{\downarrow,2}    + B_{15}\, \no_{\downarrow,1} \no_{\uparrow,2} \no_{\downarrow,2}     + B_{16}\, \no_{\uparrow,1} \no_{\downarrow,1} \no_{\uparrow,2} \no_{\downarrow,2}  .
\end{align}
The total Hamiltonian whose integrability we will investigate is
\begin{align}
\lH = \lK_{Hub} + \lK_{pair} + \lK_{flip} + V,
\end{align}
which has 22 free parameters. It is the most general Hamiltonian which preserves the number of electrons and whose single electron hopping term is given by the standard kinetic term $\lK_{Hub}$.

\paragraph{Integrable solutions}

Following our procedure, we find four integrable solutions that have an $R$-matrix of difference form. These models do not include the usual Hubbard model since that model has an $R$-matrix that cannot be written in difference form. We find that there are no integrable models of this type which have a non-zero pair hopping term $\lK_{pair}$.

First, there are three independent models that only have a non-trivial $V$
\begin{align}
\lH^{(15)} &= \mathcal{K}_{Hub}  + a_1(\no_{\uparrow,1}-\no_{\uparrow,2})^2 + a_2(\no_{\uparrow,1}-\no_{\uparrow,2})+ a_3(\no_{\downarrow,1}-\no_{\downarrow,2})^2 +
 a_4(\no_{\downarrow,1}-\no_{\downarrow,2}) \\
  \lH^{(16)} &=  \mathcal{K}_{Hub}  + a_1(\no_{\uparrow,1}-\no_{\uparrow,2})^2 + a_2(\no_{\uparrow,1}-\no_{\uparrow,2})+ a_3(\no_{\downarrow,1}+\no_{\downarrow,2}) +
 a_4(\no_{\downarrow,1}-\no_{\downarrow,2}) \\
 \lH^{(17)} &= \mathcal{K}_{Hub}  + a_1(\no_{\uparrow,1}+\no_{\uparrow,2}) + a_2(\no_{\uparrow,1}-\no_{\uparrow,2})+ a_3(\no_{\downarrow,1}+\no_{\downarrow,2}) +
 a_4(\no_{\downarrow,1}-\no_{\downarrow,2}) 
\end{align}
These models separate and the Hamiltonians can be written as
\begin{align}
\lH = \lH_\uparrow + \lH_\downarrow.
\end{align}
Hence they are simply a direct sum of two two-dimensional integrable systems. 

\paragraph{New model} Secondly, we find a Hamiltonian that has a non-trivial spin flip interaction $\lK_{flip}$ as well as a potential part
\begin{align}
\lH^{(18)} =&~ \mathcal{K}_{Hub}  +  a\Big( \cdo_{\uparrow,1}\cdo_{\downarrow,2}\co_{\downarrow,1}\co_{\uparrow,2} +  \cdo_{\downarrow,1}\cdo_{\uparrow,2}\co_{\uparrow,1}\co_{\downarrow,2}
+   \cdo_{\uparrow,1}\cdo_{\uparrow,2}\co_{\downarrow,1}\co_{\downarrow,2} + \cdo_{\downarrow,1}\cdo_{\downarrow,2}\co_{\uparrow,1}\co_{\uparrow,2}
\Big)
+\nonumber\\
&~ (2a-b) (\no_{\uparrow,1}+\no_{\downarrow,1}) +b(\no_{\uparrow,2}+\no_{\downarrow,2})  -a  (\no_{\uparrow,1}+\no_{\downarrow,1}) (\no_{\uparrow,2}+\no_{\downarrow,2})  .
\end{align}
Notice that this model does not preserve spin orientation and consequently is a type of XYZ deformation of the Hubbard potential. This model clearly does not separate as Models 15-17 did and to our knowledge is a new model of electrons on a one-dimensional lattice. The model has two free parameters and could have very interesting limits, spectral reductions and phase diagram. Since spin is not conserved in this model, the conventional Bethe Ansatz approach is not applicable. It would be a very interesting problem to find the spectrum of this model and to study its physical properties or quantum algebraic formalism derived model.

\paragraph{R-matrices} The $R$-matrix for the XYZ-type Hubbard model corresponding to $\lH^{(18)}$ is given by
\begin{equation}
R^{(18)}_{12}(u)=f(u){\tiny \left(
\begin{array}{cccccccccccccccc}
 r_1 & 0 & 0 & 0 & 0 & 0 & 0 & 0 & 0 & 0 & 0 & 0 & 0 & 0 & 0 & 0 \\
 0 & r_2 & 0 & 0 & r_6 & 0 & 0 & 0 & 0 & 0 & 0 & -r_8 & 0 & 0 & r_8 & 0 \\
 0 & 0 & r_3 & 0 & 0 & 0 & 0 & 0 & r_7 & 0 & 0 & 0 & 0 & 0 & 0 & 0 \\
 0 & 0 & 0 & r_3 & 0 & 0 & 0 & 0 & 0 & 0 & 0 & 0 & r_7 & 0 & 0 & 0 \\
 0 & r_{12} & 0 & 0 & r_2 & 0 & 0 & 0 & 0 & 0 & 0 & -r_9 & 0 & 0 & r_9 & 0 \\
 0 & 0 & 0 & 0 & 0 & r_1 & 0 & 0 & 0 & 0 & 0 & 0 & 0 & 0 & 0 & 0 \\
 0 & 0 & 0 & 0 & 0 & 0 & -r_3 & 0 & 0 & s_7 & 0 & 0 & 0 & 0 & 0 & 0 \\
 0 & 0 & 0 & 0 & 0 & 0 & 0 & -r_3 & 0 & 0 & 0 & 0 & 0 & s_7 & 0 & 0 \\
 0 & 0 & s_7 & 0 & 0 & 0 & 0 & 0 & r_3 & 0 & 0 & 0 & 0 & 0 & 0 & 0 \\
 0 & 0 & 0 & 0 & 0 & 0 & r_7 & 0 & 0 & -r_3 & 0 & 0 & 0 & 0 & 0 & 0 \\
 0 & 0 & 0 & 0 & 0 & 0 & 0 & 0 & 0 & 0 & r_4 & 0 & 0 & 0 & 0 & r_{10} \\
 0 & r_9 & 0 & 0 & r_8 & 0 & 0 & 0 & 0 & 0 & 0 & r_5 & 0 & 0 & r_{11} & 0 \\
 0 & 0 & 0 & s_7 & 0 & 0 & 0 & 0 & 0 & 0 & 0 & 0 & r_3 & 0 & 0 & 0 \\
 0 & 0 & 0 & 0 & 0 & 0 & 0 & r_7 & 0 & 0 & 0 & 0 & 0 & -r_3 & 0 & 0 \\
 0 & -r_9 & 0 & 0 & -r_8 & 0 & 0 & 0 & 0 & 0 & 0 & r_{11} & 0 & 0 & r_5 & 0 \\
 0 & 0 & 0 & 0 & 0 & 0 & 0 & 0 & 0 & 0 & r_{10} & 0 & 0 & 0 & 0 & r_4 \\
\end{array}
\right)}
\end{equation}
where we have the following functions
\begin{equation}
\begin{array}{c}
\begin{split}
& r_1=\cos (\theta +u \cos (\theta ))\\
& r_2 = \frac{\displaystyle r_3^2}{\displaystyle r_1}\\
& r_3=\sin (u \cos (\theta ))\\
& r_4=\cos (\theta ) \cos (u \cos (\theta ))\\
\end{split}
\end{array},\quad
\begin{array}{c}
\begin{split}
& r_5=-\cos (\theta ) \tan (\theta +u \cos (\theta ))r_3\\
& r_6 = \frac{\displaystyle r_7^2}{\displaystyle r_1}\\
& r_7=\cos (\theta ) e^{u (a_2+\sin (\theta ))}\\
& r_8=\frac{\displaystyle r_7}{\displaystyle r_1} r_3
\end{split}
\end{array}
\end{equation}
$$
\begin{array}{c}
\begin{split}
& r_9=\frac{\displaystyle s_7}{\displaystyle r_1} r_3\\
& r_{10}=\sin (\theta ) r_3\\
& r_{11}=\frac{1}{4} (\cos (2 \theta )-\cos (2 u \cos (\theta ))+\cos (2 (\theta +u \cos (\theta )))+3)\ r_1^{-1}\\
& r_{12}= \frac{\displaystyle s_7^2}{\displaystyle r_1}\\
\end{split}
\end{array}
$$
and have defined $s_7(u)=r_7(-u)$ and $f(u)=(2 a_2 u+1) \sec (\theta +u \cos (\theta ))$ . The $R$-matrix satisfies the braided unitarity condition
\begin{equation}
R_{12}(u)R_{21}(-u)=1-4a_2^2u^2.
\end{equation}

\section{Towards the spectrum for Models 8, 9 and 10}\label{sec:BA}

In this section we construct an ansatz for the spectrum of model 8, 9 and 10.

Models 8, 9 and 10 have $ G = H = K = L = 0 $. Therefore we deal with them at the same time. 
We succeeded to construct their eigenvalues and eigenvectors for up to two excitations for any number of sites. 
A general ansatz for any number of excitations, however, is still unknown.

\

\textbf{Notation:} In equation \eqref{phi12psi12} we see that $ |\phi_1\rangle $ is the vacuum. Therefore, we will refer to a state with $ L $ $  \phi_1 $'s as the vacuum of an L-sites spin chain. Also in equation \eqref{phi12psi12} we see that if we have a $ |\psi_1\rangle $ or a $|\psi_2\rangle $, a particle is created in that site, while if we have a $ |\phi_2\rangle $ two particles are created in that site. So, a state with a $ \psi_{\alpha} $ in one site and $ \phi_1 $'s in all other sites will be called 1-excitation. A state with two $\psi_{\alpha}  $'s or one $ \phi_2 $ and the rest $ \phi_1 $'s will be called 2-excitations, and so one and so forth. 

\subsection{Vacuum}

Let us define the vacuum as 
\begin{equation}
|\Lambda_0\rangle \equiv |\phi_1 \phi_1 ... \phi_1\rangle .
\end{equation}
According to \eqref{Hphiphi} $ \lH_{12} |\phi_1 \phi_1 \rangle = (A+B)  |\phi_1 \phi_1 \rangle $, so for the reference state, a periodic spin chain of $ L $ sites has eigenvalue
\begin{equation}
\Lambda_0=L(A+B).
\label{lambda0} 
\end{equation}
Notice that for model 8 and 9 is $ \Lambda_0=0 $ because $ A=-B $. For model 10, this does not happen.

\subsection{$ 1 $-excitation}

So, the eigenstates with $ 1 $-excitation are 
\begin{equation}
|\Lambda_{1}\rangle=|\psi_{\alpha,j}\rangle\equiv|\phi_1...\phi_1\underbrace{\psi_\alpha}_{j\text{-th site}}\phi_1...\phi_1\rangle,
\label{halfeig}
\end{equation}
where the $ \psi_\alpha $ is in the $ j$-th site of the spin chain. The corresponding eigenvalues are given by
\begin{equation}
\Lambda_{1}=(L-2)(A+B),
\label{lambdahalf}
\end{equation}
with degeneracy $ d=2L $. The reason for the degeneracy is that there are $ L $ different positions to put the $ \psi_\alpha $, and two possible values for $ \alpha $.

\subsection{$ 2 $-excitations}

For $ 2 $-excitations, as already mentioned, we can have one $ \phi_2 $ or two $ \psi_\alpha $'s. There are  $ 2L^2-L $ eigenvalues (and corresponding eigenvectors) with 2-excitations. Let us see how to construct them.

\paragraph{Case 1}

Let us start by considering the case with the two $ \psi_{\alpha} $'s being separated by one or more $ \phi_1 $'s, i.e.
\begin{equation}
|\phi_1...\phi_1\underbrace{\psi_{\alpha}}_{j\text{-th site}}\phi_1...\phi_1\underbrace{\psi_{\beta}}_{k\text{-th site}}\phi_1...\phi_1\rangle 
\label{vecsep}
\end{equation}
with $ k>j+1 $.

In this case the only contribution comes from the action of the Hamiltonian in pairs of $ \phi_1 $'s. The states constructed in this way are already eigenstates of the Hamiltonian and have eigenvalue equal to $ (L-4)(A+B)  $ for $ L>3 $. For $ L=3 $ there is no way of the eigenstate \eqref{vecsep} existing because of the periodicity of the spin chain.\

\paragraph{Case 2}
Another case considered separately is when the two $ \psi_{\alpha} $'s are together but they are equal to each other,i.e.
\begin{equation}
|\phi_1...\phi_1\underbrace{\psi_{\alpha}}_{j\text{-th site}}\underbrace{\psi_{\alpha}}_{(j+1)\text{-th site}}\phi_1...\phi_1\rangle. 
\end{equation}
Such state will be an eigenstate of the Hamiltonian with eigenvalue $ (L-3)(A+B)+(D+E)  $.

\paragraph{Case 3}
The most general state with $ 2 $-excitations (excluding case 1 and case 2) can be written as
\begin{equation}
|\Lambda_2\rangle \equiv \sum_{j}c_j|j\rangle +\sum_{j}h_{\alpha\beta,j}|\psi_{\alpha,j}\psi_{\beta,j+1}\rangle,
\label{state1exc}
\end{equation}  
where $ \alpha\neq \beta $, sum in repeated greek indices is assumed, and
\begin{align}
& |\psi_{\alpha_L}\psi_{\beta_{L+1}}\rangle=|\psi_{\beta,1}\psi_{\alpha,L}\rangle,\\
& c_0=c_L,\quad c_{L+1}=c_1,\\
& h_{\alpha\beta,0}=h_{\alpha\beta,L}, \quad h_{\alpha\beta,L+1}=h_{\alpha\beta,1},
\end{align}
and
\begin{align}
& |j\rangle \equiv |\phi_1 ...\phi_1 \underbrace{\phi_2}_{j\text{-th site}} \phi_1...\phi_1\rangle ,\\
& |\psi_{\alpha_j}\psi_{\beta_{j+1}}\rangle\equiv |\phi_1...\phi_1\underbrace{\psi_\alpha}_{j\text{-th site}}\underbrace{\psi_\beta}_{(j+1)\text{-th site}}\phi_1...\phi_1\rangle.
\end{align}
Now let us see for which values of $ c_i $ and $ h_{\alpha\beta,j} $ the state $ |\Lambda_1\rangle  $ will be an eigenstate of the Hamiltonian.

Let us start by seeing how the $ L $-sites Hamiltonian $ \mathbb{H} $ acts on the state \eqref{state1exc}. Following \eqref{Hphiphi}-\eqref{Hpsipsi} we obtain 
\begin{align}
\mathbb{H}|j\rangle=& (L-2)(A+B)|j\rangle +2A |j\rangle +B|j-1\rangle +B|j+1\rangle+\nonumber\\
& \hspace{1.2cm}+ C\epsilon^{\alpha\beta}\left(|\psi_{\alpha,j-1}\psi_{\beta,j}\rangle -|\psi_{\alpha,j}\psi_{\beta,j+1}\rangle\right),
\label{Hj}
\end{align}
and
\begin{align}
\mathbb{H}|\psi_{\alpha,j}\psi_{\beta,j+1}\rangle=& (L-3)(A+B)|\psi_{\alpha,j}\psi_{\beta,j+1}\rangle+\nonumber\\
& +D |\psi_{\alpha,j}\psi_{\beta,j+1}\rangle+E|\psi_{\beta,j}\psi_{\alpha,j+1}\rangle+F\epsilon_{\alpha\beta}\left(|j+1\rangle -|j\rangle \right).
\label{Hpsis}
\end{align}
Now we can write the action of the Hamiltonian in \eqref{state1exc} as
\begin{align}
\mathbb{H}|\Lambda_2\rangle = & \sum_{j}^{L}\left[ c_j(L-2)(A+B)+2Ac_j+Bc_{j+1}+Bc_{j-1}+\right.\nonumber\\
& \hspace{1cm}\left. + h_{\alpha\beta,j-1}\epsilon^{\alpha\beta}F-h_{\alpha\beta,j}\epsilon^{\alpha\beta}F\right]|j\rangle +\nonumber\\
& +\sum_{j=1}^{L}\left[(L-3)(A+B)h_{\alpha\beta,j}+c_{j+1}\epsilon^{\alpha\beta}C-c_j\epsilon^{\alpha\beta}C+\right.\nonumber\\
& \hspace{1cm}\left. +D h_{\alpha\beta,j}+Eh_{\beta\alpha,j}\right] |\psi_{\alpha,j}\psi_{\beta,j+1}\rangle.
\label{Hact1exc}
\end{align}
To find this formula we used the fact that the sum in $ j $  is periodic to relabel the coefficients and let everything in terms of $ |j\rangle $ and $ |\psi_{\alpha,j}\psi_{\beta,j+1}\rangle $. 

But $ \mathbb{H}|\Lambda_2\rangle  $ is also equal to 
\begin{equation}
\mathbb{H}|\Lambda_2\rangle =\sum_{j}c_j\Lambda_2|j\rangle +\sum_{j}h_{\alpha\beta,j}\Lambda_2|\psi_{\alpha,j}\psi_{\beta,j+1}\rangle.
\label{Hact1exc2}
\end{equation}
By comparing \eqref{Hact1exc} with \eqref{Hact1exc2} we obtain the two following conditions
\begin{align}
& c_j(L-2)(A+B)+2Ac_j+Bc_{j+1}+Bc_{j-1}+  h_{\alpha\beta,j-1}\epsilon^{\alpha\beta}F-h_{\alpha\beta,j}\epsilon^{\alpha\beta}F=\Lambda_2 c_j,\label{Energyc}\\
& (L-3)(A+B)h_{\alpha\beta,j}+c_{j+1}\epsilon^{\alpha\beta}C-c_j\epsilon^{\alpha\beta}C+D h_{\alpha\beta,j}+Eh_{\beta\alpha,j}=\Lambda_2 h_{\alpha\beta,j}. \label{Energyh}
\end{align}
Multiplying the equation \eqref{Energyc} by $ h_{\gamma\delta,k} $ and then using the equation \eqref{Energyh} we obtain 
\begin{align}
& (A+B)c_jh_{\gamma\delta,k}+(2A-D)c_jh_{\gamma\delta,k}+B(c_{j+1}+c_{j-1})h_{\gamma\delta,k}-Ec_jh_{\delta\gamma,k}+\nonumber\\
&\hspace{1cm}+F\epsilon^{\alpha\beta}(h_{\alpha\beta,j-1}-h_{\alpha\beta,j})h_{\gamma\delta,k}-c_j(c_{k+1}-c_k)\epsilon_{\gamma\delta}C=0.
\label{conditions}
\end{align}
Remember $ j,k=1,...,L $, $ \alpha,\beta,\gamma,\delta=1,2 $ and sum in repeated greek indices is assumed. 

The equations \eqref{conditions} are the conditions $ c_j $ and $ h_{\alpha\beta,j} $ have to satisfy in order to $ |\Lambda_2\rangle $ be eigenstates of the Hamiltonian and $ \Lambda_2 $ be its eigenvalues.

The procedure is then the following, one solves the equations \eqref{conditions} and substitutes the $ c_j $ and $ h_{\alpha\beta,j} $  found, in the equations \eqref{Energyc} and \eqref{Energyh} to obtain the eigenvalues, and substitute them on \eqref{state1exc} to obtain the eigenvectors.

All the equations presented in this subsection are valid for 2-excitations for any $ L>2 $. For $ L=3,4,5 $ we solved equations \eqref{Energyc}, \eqref{Energyh} and \eqref{conditions} algebraically (using Software Mathematica) and obtained the eigenvalues and eigenvectors. The eigenvalues obtained by this method matched completely with the ones presented in the Tables \eqref{table:model8L3}-\eqref{table:model10L5} which were computed by direct diagonalization of the Hamiltonian.

\subsection{$ 3 $-excitations and more }

For more than $ 2 $-excitations we were unable to find a general formula for the eigenvalues and eigenvectors.  For $ L=3,4,5 $ we computed the spectrum making use of the reduced Hamiltonians as presented in section \ref{sec:reducedH}. It would be very interesting to compute ABA for this models since at least model 9 and 10 seem to have very interesting physical properties.

Remember that there is a symmetry $ p\leftrightarrow (2L-p) $ with $ p $ being the number of excitations.
Therefore, since we know the eigenvalues for $ p=0 $, $ p=1 $ and $ p=2 $, we automatically have the ones for $ p=2L $, $ p=2L-1 $ and $ p=2L-2 $.

%%%%%%%%%%%%%%%%%%%%%%%%%%%%%%%%%%%%%%%%%%%%%%%%%%%%%%%%%%%%%%%%%%%%%%%%%%%%%%%%
%%%%%%%%%%%%%%%%%%%%%%%%%%%%%%%%%%%%%%%%%%%%%%%%%%%%%%%%%%%%%%%%%%%%%%%%%%%%%%%%

\section{Discussion}

In this paper we classified integral spin chains which can be identified with electrons in a conduction band. This means that we consider a four-dimensional local Hilbert space where each site can be empty, contain a single electron or an electron pair. We then use the recently proposed method of \cite{deLeeuw:2019zsi} to classify all integrable models that have additional spin and charge symmetry and whose $R$-matrix is of difference form. We recover all known models that exhibit this symmetry, such as the $\alg{su}(4)$ and $\alg{sp}(4)$ spin chain, but in addition we find several new models. 

The models that we were not able to identify are Models 4, 6, 8, 9, 10, 14 and 18. Models 4 and 6 are a slightly modified version of the $\alg{su}(4)$ and $\alg{su}(2|2)$ spin chain. They have rational $R$-matrices which hints at an underlying Yangian symmetry. It would be an interesting question to see if these models can be generalized to $\alg{su}(m)\times\alg{su}(n)$. More interesting seem to be models 8, 9 and 10, in which the fermionic degrees of freedom seem to freeze out and the only dynamical degrees of freedom correspond to electron pairs. There are many interesting research directions that can now be pursued.

First and foremost, it would be interesting to do a full classification of with less symmetric models. For instance, only assuming that spin and charge are preserved, there are 35 free parameters. One quickly finds that there will be several thousand solutions, which should contain many new and interesting integrable models. So far we have not been able to find all solutions. Many of these solutions will naturally be  related by basis transformations, twist and reparameterizations. It is very technically challenging to perform this identification. 

Secondly, it is important to find the spectrum of the new models that we have found. The new models do not seem to be solvable by means of the standard coordinate (nested) Bethe Ansatz. Using the algebraic Bethe Ansatz approach might be a way to derive the spectrum. At the very least it would be interesting to work out the RTT relations and find the quantum algebras that underly these models. 

After this, it would be interesting to study the physical properties of the new models. It would be particularly worthwhile to consider the thermodynamic or continuum limit. Indeed, Model 18 actually depends on two coupling constants and might have a non-trivial phase diagram. Moreover, it would be very interesting to investigate if there are two-dimensional field theories whose scattering matrix would correspond to the $R$-matrices of our new models.

Finally, there are some further open questions regarding our approach to finding integrable systems of different sizes. In this paper, we again confirm that $[\bQ_2,\bQ_3]=0$ is a sufficient condition for these models. It would be very interesting to see an integrable model for which $[\bQ_3,\bQ_4] = 0 $ would impose new constraints.  We can also apply our method to look at other types of models. For instance we could consider models in higher dimensions and look at generalized Hubbard models of the type \cite{Drummond:2007sa,Drummond:2007gt} or consider three dimensional models and compare with a recent paper where a set of these solutions where recently classified \cite{Vieira:2019vog}. Similarly, it would be interesting to try to generalize the construction of  \cite{1998PhLA..239..187M,1998NuPhB.517..395M} in order to obtain $\alg{su}(n)$ type Hubbard models.

\paragraph{Acknowledgements.}
We would like to thank N. Beisert, S. A. Frolov, S. Mozgovoy, R. Nepomechie, D. Volin and C. Paletta. MdL was supported by SFI, the Royal Society and the EPSRC for funding under grants UF160578, RGF$\backslash$EA$\backslash$181011, RGF$\backslash$EA$\backslash$180167 and 18/EPSRC/3590. A.P. is supported by the grant RGF$\backslash$EA$\backslash$180167. A.L.R. is supported by the grant 18/EPSRC/3590. 
 The work of P.R. is supported in part by a Nordita Visiting PhD Fellowship and by SFI and the Royal Society grant UF160578.

\appendix

\section{Grading}\label{app:Grading}
In this appendix we review the well-known formalism for associating solutions of the Yang-Baxter equation to graded solutions of the Yang-Baxter equation. We closely follow \cite{HubBook}. The ungraded $R$-matrix is defined as
\begin{equation}
R_{jj+1}(u)=R_{\beta\delta}^{\alpha\gamma}(u)E_{j\,\alpha}^{\hspace{0.20cm}\beta}E_{j+1\,\gamma}^{\hspace{0.6cm}\delta},
\label{nongradedR}
\end{equation} 
\noindent
with 
\begin{equation}
E_{j\,\alpha}^{\hspace{0.20cm}\beta}=1^{\otimes(j-1)}\otimes E_\alpha^\beta\otimes 1^{\otimes(L-j)},
\label{nongradedEj}
\end{equation}
\noindent
where $E_\alpha^\beta$ are the usual unit matricies $\left(E_\alpha^\beta\right)_i^j=\delta_i^\alpha \delta_j^\beta$ and $ \alpha,\beta=1,...,4 $.

We start by introducing the graded permutation operator defined as
\begin{equation}
P_{jj+1}^f=(-1)^{p(\beta)}e_{j\,\alpha}^{\hspace{0.2cm}\beta}e_{j+1\,\beta}^{\hspace{0.6cm}\alpha}.
\label{gradedP}
\end{equation}
where
\begin{equation}
e_{j\,\alpha}^{\hspace{0.20cm}\beta}=(-1)^{(p(\alpha)+p(\beta))\sum_{k=j+1}^{L}p(\gamma_k)}1^{\otimes (j-1)}\otimes E_\alpha^\beta\otimes E_{\gamma_{j+1}}^{\gamma_{j+1}}\otimes...\otimes E_{\gamma_{L}}^{\gamma_{L}}.
\label{gradedej}
\end{equation}
are graded local projection operators. Since we are considering a $ 2|2 $ graded vector space as our Hilbert space we therefore assume $ p(1)=p(2)=0 $ and $ p(3)=p(4)=1 $ and summation over repeated Greek indices is assumed.

The fermionic $R$-matrix $R_{jj+1}^f$ is then defined as
\begin{equation}
R_{jj+1}^f(u)=(-1)^{p(\gamma)+p(\alpha)(p(\beta)+p(\gamma))}R_{\gamma\delta}^{\alpha\beta}(u)e_{j\,\alpha}^{\hspace{0.20cm}\gamma}e_{j+1\,\beta}^{\hspace{0.6cm}\delta},
\label{gradedR}
\end{equation}
\noindent
which satisfies the Yang-Baxter equation 
\begin{equation}
R_{12}^f(u-v)R_{13}^f(u)R_{23}^f(v)=R_{23}^f(v)R_{13}^f(u)R_{12}^f(u-v),
\label{gradedYBE}
\end{equation}
provided the following compatibility condition on the entries $R_{\gamma\delta}^{\alpha\beta}$ is satisfied: 
\begin{equation}
R_{\gamma\delta}^{\alpha\beta}=(-1)^{p(\alpha)+p(\beta)+p(\gamma)+p(\delta)}R_{\gamma\delta}^{\alpha\beta}
\end{equation}
which is satisfied for all of the models considered in this paper. For $R$-matricies satisfying this constraint it is well-known that there is a one-to-one correspondence between solutions of the Yang-Baxter equation and solutions of the graded Yang-Baxter equation, see \cite{kulish1990solutions}. Note that $ R_{13}^f(u)=P_{12}^fR_{23}^f(u)P_{12}^f $ and $ P_{13}^f =P_{12}^fP_{23}^fP_{12}^f$. The fermionic $R$-matrix satisfies regularity
\begin{equation}
R_{jk}^f(0)=P_{jk}^f,
\label{gradedregularity}
\end{equation}
\noindent
and unitarity
\begin{equation}
R_{jk}^f(u)R_{kj}^f(-u)\propto 1.
\end{equation}

Using the fermionic $R$-matrix \eqref{gradedR} we are able to construct the corresponding Hamiltonian
\begin{equation}
H_{jj+1}^f=\partial_u\left(P_{jj+1}^fR_{jj+1}^f(u)\right)\Big\vert_{u=0}.
\label{gradedH}
\end{equation}

\section{4D representations of $\alg{su}(2)\times\alg{su}(2)$}\label{app:su2xsu2}\label{app:su2}

We are interested in models that have a non-trivial $\alg{su}(2)\times\alg{su}(2)$ symmetry. Since we only consider a four dimensional local Hilbert space, we need to classify all four-dimensional representations of this semi-simple Lie algebra. Clearly any representation of $\alg{su}(2)\times\alg{su}(2)$ automatically induces a four-dimensional representation on both $\alg{su}(2)$ factors.

Let us first focus on the first copy of $\alg{su}(2)$. There are five possible four-dimensional representations of $\alg{su}(2)$. Most of them are reducible and hence they decompose into irreducible representations. In particular, we find the following decompositions $1\oplus1\oplus1\oplus1, 2\oplus1\oplus1, 2\oplus2, 3\oplus1, 4$. We are only interested in non-trivial representations, hence we will not consider $1\oplus1\oplus1\oplus1$. Fixing this four-dimensional representation, we then consider three general $4\times4$ matrices and impose that they form an $\alg{su}(2)$ algebra and commute with the representation of the first $\alg{su}(2)$ factor. This is enough, up to some trivial similarity transformations, to fix the representation of the second $\alg{su}(2)$ factor. Because of this, we can denote the representation of $\alg{su}(2)\times\alg{su}(2)$ by the representation of only one factor.

Let $t^{L/R}_i$ denote the first and second set of $\alg{su}(2)$ generators in $\alg{su}(2)\times\alg{su}(2)$ respectively. They satsify
\begin{align}
[t^{L/R}_i,t^{L/R}_j] =  \epsilon_{ijk} t^{L/R}_k.
\end{align}
Let $\rho_n(t_i)$ denote the $n$-dimensional irreducible representation of the generators $t_i$ of $\alg{su}(2)$, then we find the following representations.

\paragraph{$2\oplus1\oplus1$} In this case, the two dimensional representation is embedded as a direct sum. More explicitly, the first factor is simply embedded in the upper left $2\times 2$ block, \textit{i.e.}
\begin{align}
\rho_{2\oplus1\oplus1} (t^L_i) = \rho_2(t_i) \oplus 0=
\begin{pmatrix}
\rho_2(t_i) & 0\\
0 & 0
\end{pmatrix}.
\end{align}
It is easy to see that this immediately implies that the second representation has to be two-dimensional and embedded in the lower right block
\begin{align}
\rho_{2\oplus1\oplus1} (t^R_i) = 0\oplus \rho_2(t_i) = 
\begin{pmatrix}
0& 0\\
0 & \rho_2(t_i) 
\end{pmatrix}.
\end{align}
These are the usual spin and charge $\alg{su}(2)$ symmetries of the Hubbard model.

\paragraph{$2\oplus2$} In this situation, the two-dimensional $\alg{su}(2)$ representation is embedded diagonally and can be written as 
\begin{align}
\rho_{2\oplus2} (t^L_i)  = 1 \otimes  \rho_2(t_i) = \begin{pmatrix}
\rho_2(t_i) & 0\\
0 & \rho_2(t_i) 
\end{pmatrix}.
\end{align}
It is easy to check that the only other non-trivial representation of $\alg{su}(2)$ commuting with this is 
\begin{align}
\rho_{2\oplus2} (t^R_i)  =  \rho_2(t_i) \otimes 1.
\end{align}
This representation is isomorphic to $\alg{so}(4)$.

\paragraph{$3\oplus1$ and $4$} When one of the $\alg{su}(2)$ representations contains a three- or four-dimensional irreducible component, then it is straightforward to show that there is no non-trivial representation for the second $\alg{su}(2)$ representation that commutes with it. Because of this we will not consider these representations since they would amount to just considering a $\alg{su}(2)$ type spin chain.

\section{The spectrum for models 8, 9 and 10}\label{sec:reducedH}

The main objective of this appendix is to present the eigenvalues and degeneracies computed by direct diagonalization of the Hamiltonians for models 8, 9 and 10 with length $ L=3,4,5 $. The results presented below make possible to compare the degeneracies presented in section 6 with the ones obtained by numerical calculations.

This appendix is divided into two subsections. In the first we explain how we computed the eigenvalues, while in the second we put the tables with eigenvalues and degeneracies.

\subsection{Reduced Hamiltonians}

The size of the Hamiltonian increases exponentially with the number of sites, so it is not easy to directly diagonalize it. Actually, even when the direct diagonalization is possible, it is not easy to know from which type of excitation each eigenvalue is coming from, since Mathematica mixes different excitations when computing the eigenvectors. 

A better way to do this is to use reduced Hamiltonians. The idea is to construct a set $ v={v_1,...,v_m} $ of all the possible vectors with a certain number of excitations. And then if these vectors satisfy

\begin{equation}
v_i^T.v_j=\delta_{ij},
\label{orthogonality}
\end{equation}  
\noindent
with $ T $ denoting transposition, we can define the reduced Hamiltonian as 

\begin{equation}
\mathbb{H}_{red}=v_i^T.\mathbb{H}.v_j,
\label{reducedH}
\end{equation}
\noindent 
where $ i$ and $ j $ go from $ 1 $ to the total number of vectors for that number of excitations.

Let us see how this works for $ 2 $-excitations, for example. For 2-excitations we can have one $ \phi_2 $ or two $ \phi_1 $'s. So, the number of ways to put $ \phi_2 $ in a spin chain of length L is 

\begin{equation}
\text{\# of Permutations}\left(|\phi_1...\phi_1\phi_2\phi_1...\phi_1\rangle\right)=\frac{L!}{1!(L-1)!}=L, 
\end{equation}
\noindent
while the number of ways to put two $ \psi $'s is 

\[
\text{\# of Permutations}\left(|\phi_1...\phi_1\psi_\alpha\psi_\beta\phi_1...\phi_1\rangle\right)=\left\{
\begin{array}{ll}
2\left(\frac{L!}{2!(L-2)!}\right)=L(L-1)\quad\text{for}\quad \alpha=\beta\\
\left(\frac{L!}{1!1!(L-2)!}\right)=L(L-1)\quad\text{for}\quad \alpha\neq\beta
\end{array}
\right.
\]

So, for 2-excitations we have a total of $ 2L^2-L $ possible vectors, so $ i $ and $ j $ in equations \eqref{orthogonality} and $ \eqref{reducedH} $ are $ i,j=1,...,(2L^2-L) $ .
Now it is just to use these vectors to construct the reduced Hamiltonian $ \mathbb{H}_{red} $ in equation \eqref{reducedH}. By diagonalizing this Hamiltonian one obtains all the eigenvalues with 2-excitation. One can repeat this procedure until have all the possible excitations for a given number of sites.

Notice, that diagonalize $ \mathbb{H}_{red} $ is a lot easier than diagonalize the full $ \mathbb{H} $ because the size of the matrix is much smaller, and it has the advantage of providing the information of which eigenvalues come from which excitation. 

\newpage 

\subsection{Spectrum computed using the reduced Hamiltonians}

In this subsection we construct tables with the spectrum of Hamiltonians for $ L=3,4,5 $ for Models 8, 9 and 10 using the reduced Hamiltonians. 

The tables do not show all the possible excitations, because the Hamiltonian $ \mathbb{H} $ has a symmetry $ p\leftrightarrow 2L-p $ where $ p $ is the number of excitations\footnote{We checked this claim for models 8, 9 and 10 with $ L=3,4,5 $ but it is a direct consequence of the $ \alg{su}(2)\times \alg{su}(2) $ symmetry.}. It is therefore enough to show half of the excitations. For  $ L=3 $,  for example, the state with 2-excitations ($ p=2 $) has the same eigenvalues as the state with 4-excitations ($ p=4 $).

In the following, $ d $ denotes degeneracy and $ \Lambda $ denotes eigenvalue.

\subsubsection{Model 8}

For model 8, the eigenvalues for the Hamiltonians with L=2,3 4 are presented in the tables \ref{table:model8L3} - \ref{table:model8L5}

\begin{table}[h!]
	\begin{center}
		\begin{tabular}{{|c|c|}}
			\hline
			Number of excitations & d$ (\Lambda) $\\
			\hline\hline
			0 &  $ \{1(0)\} $\\
			\hline
			1&  $ \{6(0)\} $\\
			\hline
			2 & $ \{1(-2),2(1),12(0)\} $ \\ 
			\hline
			3 &  $ \{2(2),4(-1),14(0)\} $\\ 
			\hline
		\end{tabular}
	\end{center}
	\caption{Eigenvalues and their corresponding number of excitations for a chain with 3 sites for model 8. }
	\label{table:model8L3}
\end{table}

\begin{table}[h!] 
	\begin{center}
		\begin{tabular}{{|c|c|}}
			\hline
			Number of excitations &  d $ (\Lambda) $\\
			\hline\hline
			0 &  $ \{1(0)\} $\\
			\hline
			1 &  $ \{8(0)\} $\\
			\hline
			2 &  $ \{1(-2),1(2),26(0)\}$ \\ 
			\hline
			3 & $ \{4(-2),4(2),48(0)\} $\\ 
			\hline
			\multirow{2}{0.8em}{4} & $ \left\{1(2i),1(-2i),4(-2),4(2),58(0),\right. $\\
			& $ \left.1(-2\sqrt{3}),1(2\sqrt{3}) \right\} $\\
			\hline
		\end{tabular}
	\end{center}
	\caption{Eigenvalues and their corresponding number of excitations for a chain with 4 sites  for model 8. }
\label{table:model8L4}
\end{table}

\begin{table}[h!]
	\begin{center}
		\begin{tabular}{{|c|c|}}
			\hline
			Number of excitations & d $ (\Lambda) $\\
			\hline\hline
			0 & $\{1(0)\}$\\
			\hline
			1 & $\{10(0)\}$\\
			\hline
			2 &  $\left\{2 \left( \frac{1}{2}(1+\sqrt{5}) \right),2 \left( \frac{1}{2}(1-\sqrt{5}) \right),1(-2),40(0) \right\}$\\ 
			\hline
			\multirow{2}{0.8em}{3} &  $\left\{4 \left( \frac{1}{2}(-1+\sqrt{5}) \right),4 \left( \frac{1}{2}(-1-\sqrt{5}) \right),4(-1),2(2),90(0),\right.$\\
			& $ \left. 4(-1.9563),4(1.82709),4(1.33826),4(-0.209057)\right\}  $\\
			\hline
			\multirow{3}{0.8em}{4} &  $\left\{8 \left( \frac{1}{2}(1+\sqrt{5}) \right),8 \left( \frac{1}{2}(1-\sqrt{5}) \right), 1(-1+\sqrt{5}),1(-1-\sqrt{5}),\right.$\\
			& $ 2\left(\frac{1}{2}(3+\sqrt{5})\right),2\left(\frac{1}{2}(3-\sqrt{5})\right),4(-2),4(-1),8(1),140(0),  $\\
			& $ \left.8(1.9563),8(-1.82709),8(-1.33826),8(0.209057)\right\} $\\
			\hline
			\multirow{3}{0.8em}{5} &  $\left\{12 \left( \frac{1}{2}(-1+\sqrt{5}) \right),12 \left( \frac{1}{2}(-1-\sqrt{5}) \right),4 ( \frac{1}{2}(-3+\sqrt{5}) ),\right.$\\
			& $  4 \left( \frac{1}{2}(-3-\sqrt{5}) \right),2(1+\sqrt{5}),2(1-\sqrt{5}),6(2),8(-1),8(1), $\\
			& $ \left.162(0),8(-1.9563),8(1.82709),8(1.33826),8(-0.209057)\right\} $\\
			\hline
		\end{tabular}
	\end{center}
	\caption{Eigenvalues and their corresponding number of excitations for a chain with 5 sites for model 8. }
\label{table:model8L5}
\end{table}

\subsubsection{Model 9 }

For model 9, the eigenvalues and degeneracies are presented in Tables \ref{table:model9L3} - \ref{table:model9L5}. 

\begin{table}[h!]
	\begin{center}
	\begin{tabular}{{|c|c|}}
		\hline
		Number of excitations & d $ (\Lambda) $\\
		\hline\hline
		0 &  $ \{1(0)\} $\\
		\hline
		1 &  $ \{6(0)\} $\\
		\hline
		2 & $ \{2(5),1(2),12(0)\} $ \\ 
		\hline
		3 &  $ \{4(5),2(2),14(0)\} $\\ 
		\hline
	\end{tabular}
	\end{center}
\caption{Eigenvalues for a chain with 3 sites for model 9. }
\label{table:model9L3}
\end{table}

\begin{table}[h!]
\begin{center}
	\begin{tabular}{{|c|c|}}
		\hline
		Number of excitations &  d $ (\Lambda) $\\
		\hline\hline
		0 &  $ \{1(0)\} $\\
		\hline
		1 &  $ \{8(0)\} $\\
		\hline
		2 &  $ \{1(6),2(4),1(2),24(0)\}$ \\ 
		\hline
		3 & $ \{4(6),8(4),4(2),40(0)\} $\\ 
		\hline
		4 & $ \{1(10),6(6),10(4),5(2),48(0)\} $\\ 
		\hline
	\end{tabular}
\end{center}
\caption{Eigenvalues and their corresponding number of excitations for a chain with 4 sites for model 9. }
\label{table:model9L4}
\end{table}

\begin{table}[h!]
\begin{center}
	\begin{tabular}{{|c|c|}}
		\hline
		Number of excitations & d $ (\Lambda) $\\
		\hline\hline
		0 & $\{1(0)\}$\\
		\hline
		1 & $\{10(0)\}$\\
		\hline
		2 &  $\left\{2 \left( \frac{1}{2}(9+\sqrt{5}) \right),2 \left( \frac{1}{2}(9-\sqrt{5}) \right),1(2),40(0) \right\}$\\ 
		\hline
		\multirow{2}{0.8em}{3} &  $\left\{4 \left( \frac{1}{2}(9+\sqrt{5}) \right),4 \left( \frac{1}{2}(9-\sqrt{5}) \right),4(5),2(2),90(0)\right.$\\
		& $ \left. 4(5.95630),4(4.20906),4(2.66174),4(2.17291)\right\}  $\\
		\hline
		\multirow{4}{0.8em}{4} &  $\left\{8 \left( \frac{1}{2}(9+\sqrt{5}) \right),8 \left( \frac{1}{2}(9-\sqrt{5}) \right),1(5+\sqrt{5}),1(5-\sqrt{5}),\right.$\\
		& $  8(5),4(2),140(0),2(10.0507),2(7.49137),  $\\
		& $ 8(5.95630),8(4.20906),2(3.8906),2(3.56732), $\\
		& $ \left.8(2.66174),8(2.17291) \right\} $\\
		\hline
		\multirow{4}{0.8em}{5} &  $\left\{12 \left( \frac{1}{2}(9+\sqrt{5}) \right),12 \left( \frac{1}{2}(9-\sqrt{5}) \right),2(5+\sqrt{5}),2(5-\sqrt{5}),\right.$\\
		& $  8(5),6(2),162(0),4(10.0507),4(7.49137),  $\\
		& $ 8(5.95630),8(4.20906),4(3.8906),4(3.56732), $\\
		& $ \left.8(2.66174),8(2.17291) \right\} $\\
		\hline
	\end{tabular}
\end{center}
\caption{Eigenvalues and their corresponding number of excitations for a chain with 5 sites for model 9. }
\label{table:model9L5}
\end{table}

\newpage
\subsubsection{Model 10}

For model 10, the eigenvalues for $ L=3,4,5 $ are presented in Tables \ref{table:model10L3} - \ref{table:model10L5}.

\begin{table}[h!]
	\begin{center}
		\begin{tabular}{{|c|c|}}
			\hline
			Number of excitations & d $ (\Lambda) $\\
			\hline\hline
			0 &  $ \{1(\frac{9}{4})\} $\\
			\hline
			1 &  $ \{6(\frac{3}{4})\} $\\
			\hline
			2 & $ \{2(\frac{23}{4}),1(\frac{11}{4}),3(\frac{9}{4}),9(\frac{3}{4})\} $ \\ 
			\hline
			3 &  $ \{4(\frac{23}{4}),2(\frac{11}{4}),8(\frac{9}{4}),6(\frac{3}{4})\} $\\ 
			\hline
		\end{tabular}
	\end{center}
	\caption{Eigenvalues and their corresponding number of excitations for a chain with 3 sites for model 10. }
	\label{table:model10L3}
\end{table}

\begin{table}[h!] 
	\begin{center}
		\begin{tabular}{{|c|c|}}
			\hline
			Number of excitations &  d $ (\Lambda) $\\
			\hline\hline
			0 &  $ \{1(3)\} $\\
			\hline
			1 &  $ \{8(\frac{3}{2})\} $\\
			\hline
			2 &  $ \{1\left(\frac{15}{2}\right),2\left(\frac{11}{2}\right),1(\frac{7}{2}),4(3),12\left(\frac{3}{2}\right),8(0)\}$ \\ 
			\hline
			\multirow{2}{0.8em}{3} & $ \left\{4(6),8(3),4(2),24\left(\frac{3}{2}\right),\right. $\\ 
			& $\left. 8\left(\frac{1}{2}(7+\sqrt{7})\right),8\left(\frac{1}{2}(7-\sqrt{7})\right) \right\} $\\
			\hline
			\multirow{3}{0.8em}{4} & $ \left\{1(9),4\left(\frac{15}{2}\right),10\left(\frac{11}{2}\right),1(5),4\left(\frac{7}{2}\right),18(3),\right. $\\
			& $ 12\left(\frac{3}{2}\right),16(0),1\left(\frac{1}{4}(29+\sqrt{145})\right),1\left(\frac{1}{4}(29-\sqrt{145})\right), $\\
			& $ \left.1\left(\frac{1}{4}(21+\sqrt{33})\right),1\left(\frac{1}{4}(21-\sqrt{33})\right) \right\} $\\
			\hline
		\end{tabular}
	\end{center}
	\caption{Eigenvalues and their corresponding number of excitations for a chain with 4 sites for model 10. }
	\label{table:model10L4}
\end{table}

\begin{table}[h!]
	\begin{center}
		\begin{tabular}{{|c|c|}}
			\hline
			Number of excitations &  d $ (\Lambda) $\\
			\hline\hline
			0 &  $ \{1\left(\frac{15}{4}\right)\} $\\
			\hline
			1 &  $ \{10(\frac{9}{4})\} $\\
			\hline
			\multirow{2}{0.8em}{2} &  $ \left\{1\left(\frac{17}{4}\right),5(\frac{15}{4}),15(\frac{9}{4}),20(\frac{3}{4}),\right.$ \\ 
			& $ \left. 2\left(\frac{1}{4}(27+2\sqrt{5})\right),2\left(\frac{1}{4}(27-2\sqrt{5})\right) \right\} $\\
			\hline
			\multirow{3}{0.8em}{3} & $ \left\{2\left(\frac{11}{4}\right),40\left(\frac{9}{4}\right),30\left(\frac{3}{4}\right),4\left(\frac{1}{4}(19+2\sqrt{7})\right), 4\left(\frac{1}{4}(19-2\sqrt{7})\right),\right. $\\ 
			& $4(6.72073),4(6.49331),4(5.52133),4(4.91537),4(4.33029), $\\
			& $\left.4(3.82412),4(3.14143),4(2.95068),4(2.83293),4(2.7698)\right\} $\\
			\hline
			\multirow{7}{0.8em}{4} & $ \left\{1\left(\frac{1}{4}(26+\sqrt{65})\right),1\left(\frac{1}{4}(26-\sqrt{65})\right),2\left(\frac{1}{4}(27+2\sqrt{5})\right),\right. $\\
			& $ 2\left(\frac{1}{4}(27-2\sqrt{5})\right),6\left(\frac{1}{4}(19+2\sqrt{7})\right),6\left(\frac{1}{4}(19-2\sqrt{7})\right), $\\
			& $2\left(\frac{29}{4}\right), 1\left(\frac{17}{4}\right), 10\left(\frac{15}{4}\right),3\left(\frac{11}{4}\right),60\left(\frac{9}{4}\right),40\left(\frac{3}{4}\right), $\\
			& $ 2 (10.9277), 2 (8.31678), 2 (8.2063), 6 (6.72073), 6 (6.49331), $\\
			& $ 2 (6.45906), 2 (6.06519), 2 (5.69032), 6 (5.52133), 6 (4.91537), $\\
			& $ 2 (4.91174), 2 (4.42291), 6 (4.33029), 6 (3.82412), 6 (3.14143), $\\
			& $ \left.6(2.95068),6(2.83293),6(2.7698)\right\} $\\
			\hline
			\multirow{7}{0.8em}{5} & $ \left\{2\left(\frac{1}{4}(26+\sqrt{65})\right),2\left(\frac{1}{4}(26-\sqrt{65})\right),8\left(\frac{1}{4}(27+2\sqrt{5})\right),\right. $\\
			& $ 8\left(\frac{1}{4}(27-2\sqrt{5})\right),4\left(\frac{1}{4}(19+2\sqrt{7})\right),4\left(\frac{1}{4}(19-2\sqrt{7})\right), $\\
			& $4\left(\frac{29}{4}\right), 4\left(\frac{17}{4}\right), 32\left(\frac{15}{4}\right),2\left(\frac{11}{4}\right),50\left(\frac{9}{4}\right),60\left(\frac{3}{4}\right), $\\
			& $ 4 (10.9277), 4 (8.31678), 4 (8.2063), 4 (6.72073), 4 (6.49331), $\\
			& $ 4 (6.45906), 4 (6.06519), 4 (5.69032), 4 (5.52133), 4 (4.91537), $\\
			& $ 4 (4.91174), 4 (4.42291), 4 (4.33029), 4 (3.82412), 4 (3.14143), $\\
			& $ \left.4(2.95068),4(2.83293),4(2.7698)\right\} $\\
			\hline
		\end{tabular}
	\end{center}
	\caption{Eigenvalues and their corresponding number of excitations for a chain with 5 sites for model 10. }
	\label{table:model10L5}
\end{table}

\newpage

\begin{bibtex}[\jobname]

@article{deLeeuw:2015ula,
      author         = "de Leeuw, Marius and Regelskis, Vidas",
      title          = "{An algebraic approach to the Hubbard model}",
      journal        = "Phys. Lett.",
      volume         = "A380",
      year           = "2016",
      pages          = "645-653",
      doi            = "10.1016/j.physleta.2015.12.013",
      eprint         = "1509.06205",
      archivePrefix  = "arXiv",
      primaryClass   = "math-ph",
      reportNumber   = "DMUS-MP-15-13",
      SLACcitation   = "%%CITATION = ARXIV:1509.06205;%%"
}

@book{HubBook, place={Cambridge}, title={The One-Dimensional Hubbard Model}, DOI={10.1017/CBO9780511534843}, publisher={Cambridge University Press}, author={Essler, Fabian H. L. and Frahm, Holger and G\"ohmann, Frank and Kl\"umper, Andreas and Korepin, Vladimir E.}, year={2005}}

@article{Tetelman,
      author         = "Tetelman, M.G.",
      title          = "{ Lorentz group for two-dimensional integrable lattice systems.}",
      journal        = "Sov. Phys. JETP",
      volume         = "55(2)",
      year           = "1982",
      pages          = "306-310",
      doi            = "",
      reportNumber   = "2",
      SLACcitation   = ""
}

@article{Kulish:1979cr,
	author         = "Kulish, P. P. and Reshetikhin, N. {\relax Yu}.",
	title          = "{Generalised Heisenberg Ferromagnet and the Gross-Neveu model}",
	journal        = "Sov. Phys. JETP",
	volume         = "53",
	year           = "1981",
	pages          = "108-114",
	note           = "[Zh. Eksp. Teor. Fiz.80,214(1981)]",
	SLACcitation   = "%%CITATION = SPHJA,53,108;%%"
}

@article{Reshetikhin:1986vd,
	author         = "Reshetikhin, N. {\relax Yu}.",
	title          = "{Integrable Models of Quantum One-dimensional Magnets
		With O($N$) and Sp(2k) Symmetry}",
	journal        = "Theor. Math. Phys.",
	volume         = "63",
	year           = "1985",
	pages          = "555-569",
	doi            = "10.1007/BF01017501",
	note           = "[Teor. Mat. Fiz.63,347(1985)]",
	SLACcitation   = "%%CITATION = TMPHA,63,555;%%"
}

@article{VGDrinfeldCQS1983,
	title={Constant quasi-classical solutions of the Yang--Baxter quantum equation, 1983},
	author={Drinfeld, V. G.},
	journal        = "Dokl. Akad, Nauk",
	volume={273},
	year           = "1983",
	pages={531-535}
}

@article{PhysRevB.12.3795,
	title = {Model for a multicomponent quantum system},
	author = {Sutherland, Bill},
	journal = {Phys. Rev. B},
	volume = {12},
	issue = {9},
	pages = {3795--3805},
	numpages = {0},
	year = {1975},
	month = {Nov},
	publisher = {American Physical Society},
	doi = {10.1103/PhysRevB.12.3795},
	url = {https://link.aps.org/doi/10.1103/PhysRevB.12.3795}
}

@article{BERG1978125,
	title = "Factorized U(n) symmetric S-matrices in two dimensions",
	journal = "Nuclear Physics B",
	volume = "134",
	number = "1",
	pages = "125 - 132",
	year = "1978",
	issn = "0550-3213",
	doi = "https://doi.org/10.1016/0550-3213(78)90489-3",
	url = "http://www.sciencedirect.com/science/article/pii/0550321378904893",
	author = "B. Berg and M. Karowski and P. Weisz and V. Kurak",
	abstract = "S-matrices describing the scattering of solitons belonging to the fundamental representation of U(n) are classified."
}

@article{10.1143/PTP.70.1508,
	author = {Fuchssteiner, Benno},
	title = "{Mastersymmetries, Higher Order Time-Dependent Symmetries and Conserved Densities of Nonlinear Evolution Equations}",
	journal = {Progress of Theoretical Physics},
	volume = {70},
	number = {6},
	pages = {1508-1522},
	year = {1983},
	month = {12},
	abstract = "{As examples, for the Lie algebras of mastersymmetries, all time-dependent symmetries and constants of motion of the Benjamin-Ono equation, the Kadomtsev-Petviashvili equation, and all their generalizations are explicitly constructed. It is shown that these quantities exist in any polynomial order of time, that they are not in involution and that they do not coincide for different members of the hierarchies. It turns out that the corresponding Lie algebras are finitely generated and that the crucial role in this generating-process is played by vector fields which are constant on the manifold under consideration. The general method for the construction of the relevant quantities is described in detail, so that it can be applied to other nonlinear evolution equations as well.}",
	issn = {0033-068X},
	doi = {10.1143/PTP.70.1508},
	url = {https://doi.org/10.1143/PTP.70.1508},
%	eprint = {http://oup.prod.sis.lan/ptp/article-pdf/70/6/1508/5208894/70-6-1508.pdf},
}

@article{Tu_2011,
	title={Effective Field Theory for theSO(n)Bilinear-Biquadratic Spin Chain},
	volume={107},
	ISSN={1079-7114},
	url={http://dx.doi.org/10.1103/PhysRevLett.107.077204},
	DOI={10.1103/physrevlett.107.077204},
	number={7},
	journal={Physical Review Letters},
	publisher={American Physical Society (APS)},
	author={Tu, Hong-Hao and Orús, Román},
	year={2011},
	month={Aug}
}

@article{Tu_2008,
	title={Class of exactly solvableSO(n)symmetric spin chains with matrix product ground states},
	volume={78},
	ISSN={1550-235X},
	url={http://dx.doi.org/10.1103/PhysRevB.78.094404},
	DOI={10.1103/physrevb.78.094404},
	number={9},
	journal={Physical Review B},
	publisher={American Physical Society (APS)},
	author={Tu, Hong-Hao and Zhang, Guang-Ming and Xiang, Tao},
	year={2008},
	month={Sep}
}

@article{Loebbert:2016cdm,
      author         = "Loebbert, Florian",
      title          = "{Lectures on Yangian Symmetry}",
      journal        = "J. Phys.",
      volume         = "A49",
      year           = "2016",
      number         = "32",
      pages          = "323002",
      doi            = "10.1088/1751-8113/49/32/323002",
      eprint         = "1606.02947",
      archivePrefix  = "arXiv",
      primaryClass   = "hep-th",
      reportNumber   = "HU-EP-16-12",
      SLACcitation   = "%%CITATION = ARXIV:1606.02947;%%"
}

@article{Vieira,
      author         = "Vieira, R. S.",
      title          = "{Solving and classifying the solutions of the Yang-Baxter
                        equation through a differential approach. Two-state
                        systems}",
      journal        = "JHEP",
      volume         = "10",
      year           = "2018",
      pages          = "110",
      doi            = "10.1007/JHEP10(2018)110",
      eprint         = "1712.02341",
      archivePrefix  = "arXiv",
      primaryClass   = "nlin.SI"
}

@article{BFdLL2013integrable,
	title={Integrable deformations of the XXZ spin chain},
	author={Beisert, Niklas and Fi{\'e}vet, Lucas and de Leeuw, Marius and Loebbert, Florian},
	journal={Journal of Statistical Mechanics: Theory and Experiment},
	volume={2013},
	number={09},
	pages={P09028},
	year={2013},
	publisher={IOP Publishing}
}

@book{Marshakov1999SeibergWitten,
	title={Seiberg-Witten theory and integrable systems},
	author={Marshakov, Andrei},
	year={1999},
	publisher={World Scientific}
}

@article{Ipsen:2018fmu,
      author         = "Ipsen, Asger C. and Staudacher, Matthias and Zippelius,
                        Leonard",
      title          = "{The one-loop spectral problem of strongly twisted $
                        \mathcal{N} $ = 4 Super Yang-Mills theory}",
      journal        = "JHEP",
      volume         = "04",
      year           = "2019",
      pages          = "044",
      doi            = "10.1007/JHEP04(2019)044",
      eprint         = "1812.08794",
      archivePrefix  = "arXiv",
      primaryClass   = "hep-th",
      reportNumber   = "HU-Mathematik-2018-11, HU-EP-18/39",
      SLACcitation   = "%%CITATION = ARXIV:1812.08794;%%"
}

@article{Gainutdinov:2016pxy,
      author         = "Gainutdinov, Azat M. and Nepomechie, Rafael I.",
      title          = "{Algebraic Bethe ansatz for the quantum group invariant
                        open XXZ chain at roots of unity}",
      journal        = "Nucl. Phys.",
      volume         = "B909",
      year           = "2016",
      pages          = "796-839",
      doi            = "10.1016/j.nuclphysb.2016.06.007",
      eprint         = "1603.09249",
      archivePrefix  = "arXiv",
      primaryClass   = "math-ph",
      reportNumber   = "UMTG-283",
      SLACcitation   = "%%CITATION = ARXIV:1603.09249;%%"
}

@article{Caux,
	year = 2011,
	month = {feb},
	publisher = {{IOP} Publishing},
	volume = {2011},
	number = {02},
	pages = {P02023},
	author = {Jean-Sebastien Caux and Jorn Mossel},
	title = {Remarks on the notion of quantum integrability},
	journal = {Journal of Statistical Mechanics: Theory and Experiment},
}

@article{Grabowski,
	year = 1995,
	month = {sep},
	publisher = {{IOP} Publishing},
	volume = {28},
	number = {17},
	pages = {4777--4798},
	author = {M P Grabowski and P Mathieu},
	title = {Integrability test for spin chains},
	journal = {Journal of Physics A: Mathematical and General},
}

@Article{Kulish1982,
author="Kulish, P. P.
and Sklyanin, E. K.",
title="Solutions of the Yang-Baxter equation",
journal="Journal of Soviet Mathematics",
year="1982",
month="Jul",
day="01",
volume="19",
number="5",
pages="1596--1620",
}

@article{Fontanella:2017rvu,
      author         = "Fontanella, Andrea and Torrielli, Alessandro",
      title          = "{Massless $AdS_2$ scattering and Bethe ansatz}",
      journal        = "JHEP",
      volume         = "09",
      year           = "2017",
      pages          = "075",
      doi            = "10.1007/JHEP09(2017)075",
      eprint         = "1706.02634",
      archivePrefix  = "arXiv",
      primaryClass   = "hep-th",
      reportNumber   = "DMUS-MP-17-05",
      SLACcitation   = "%%CITATION = ARXIV:1706.02634;%%"
}
@Article{2002quant.ph.11050D,
author="Dye, H. A.",
title="Unitary Solutions to the Yang--Baxter Equation in Dimension Four",
journal="Quantum Information Processing",
year="2003",
month="Apr",
day="01",
volume="2",
number="1",
pages="117--152",
abstract="In this paper, we determine all unitary solutions to the Yang--Baxter equation in dimension four. Quantum computation motivates this study. This set of solutions will assist in clarifying the relationship between quantum entanglement and topological entanglement. We present a variety of facts about the Yang--Baxter equation for the reader unfamiliar with the equation.",
issn="1573-1332",
doi="10.1023/A:1025843426102",
url="https://doi.org/10.1023/A:1025843426102"
}

@article{Hietarinta:1992ix,
      author         = "Hietarinta, Jarmo",
      title          = "{Solving the two-dimensional constant quantum Yang-Baxter
                        equation}",
      journal        = "J. Math. Phys.",
      volume         = "34",
      year           = "1993",
      pages          = "1725-1756",
      doi            = "10.1063/1.530185",
      reportNumber   = "TURKU-FL-R7",
      SLACcitation   = "%%CITATION = JMAPA,34,1725;%%"
}

@ARTICLE{2018arXiv180608400P,
       author = {{Pourkia}, Arash},
        title = "{Solutions to the constant Yang-Baxter equation in all dimensions}",
      journal = {arXiv e-prints},
     keywords = {Quantum Physics},
         year = "2018",
        month = "Jun",
          eid = {arXiv:1806.08400},
        pages = {arXiv:1806.08400},
archivePrefix = {arXiv},
       eprint = {1806.08400},
 primaryClass = {quant-ph},
       adsurl = {https://ui.adsabs.harvard.edu/abs/2018arXiv180608400P},
      adsnote = {Provided by the SAO/NASA Astrophysics Data System}
}

@article{Bombardelli:2018jkj,
      author         = "Bombardelli, Diego and Stefanski, Bogdan and Torrielli,
                        Alessandro",
      title          = "{The low-energy limit of AdS$_{3}$/CFT$_{2}$ and its
                        TBA}",
      journal        = "JHEP",
      volume         = "10",
      year           = "2018",
      pages          = "177",
      doi            = "10.1007/JHEP10(2018)177",
      eprint         = "1807.07775",
      archivePrefix  = "arXiv",
      primaryClass   = "hep-th",
      reportNumber   = "DMUS-MP-18-04",
      SLACcitation   = "%%CITATION = ARXIV:1807.07775;%%"
}

@article{Gromov:2017cja,
      author         = "Gromov, Nikolay and Kazakov, Vladimir and Korchemsky,
                        Gregory and Negro, Stefano and Sizov, Grigory",
      title          = "{Integrability of Conformal Fishnet Theory}",
      journal        = "JHEP",
      volume         = "01",
      year           = "2018",
      pages          = "095",
      doi            = "10.1007/JHEP01(2018)095",
      eprint         = "1706.04167",
      archivePrefix  = "arXiv",
      primaryClass   = "hep-th",
      SLACcitation   = "%%CITATION = ARXIV:1706.04167;%%"
}

@article{Caetano:2016ydc,
      author         = "Caetano, Joao and Gurdogan, Omer and Kazakov,
                        Vladimir",
      title          = "{Chiral limit of $ \mathcal{N} $ = 4 SYM and ABJM and
                        integrable Feynman graphs}",
      journal        = "JHEP",
      volume         = "03",
      year           = "2018",
      pages          = "077",
      doi            = "10.1007/JHEP03(2018)077",
      eprint         = "1612.05895",
      archivePrefix  = "arXiv",
      primaryClass   = "hep-th",
      SLACcitation   = "%%CITATION = ARXIV:1612.05895;%%"
}

@article{Beisert:2005tm,
      author         = "Beisert, Niklas",
      title          = "{The SU(2|2) dynamic S-matrix}",
      journal        = "Adv. Theor. Math. Phys.",
      volume         = "12",
      year           = "2008",
      pages          = "945-979",
      doi            = "10.4310/ATMP.2008.v12.n5.a1",
      eprint         = "hep-th/0511082",
      archivePrefix  = "arXiv",
      primaryClass   = "hep-th",
      reportNumber   = "PUTP-2181, NSF-KITP-05-92",
      SLACcitation   = "%%CITATION = HEP-TH/0511082;%%"
}

@article{Jimbo1986,
	author="Jimbo, Michio",
	title="Quantum R matrix for the generalized Toda system",
	journal="Communications in Mathematical Physics",
	year="1986",
	month="Dec",
	day="01",
	volume="102",
	number="4",
	pages="537--547",
	issn="1432-0916",
	doi="10.1007/BF01221646",
	url="https://doi.org/10.1007/BF01221646"
}

@article{Bazhanov1987,
	author="Bazhanov, V. V.",
	title="Integrable quantum systems and classical Lie algebras",
	journal="Communications in Mathematical Physics",
	year="1987",
	month="Sep",
	day="01",
	volume="113",
	number="3",
	pages="471--503",
	doi="10.1007/BF01221256",
	url="https://doi.org/10.1007/BF01221256"
}

@article{Kuniba:1991yd,
	author         = "Kuniba, Atsuo",
	title          = "{Exact solutions of solid on solid models for twisted
		affine Lie algebras $A^{(2)}_{2n}$ and $A^{(2)}_{2n-1}$}",
	journal        = "Nucl. Phys.",
	volume         = "B355",
	year           = "1991",
	pages          = "801-821",
	doi            = "10.1016/0550-3213(91)90495-J",
	reportNumber   = "SMS-073-90",
	SLACcitation   = "%%CITATION = NUPHA,B355,801;%%"
}

@article{Stolin1997,
	author="Stolin, Alexander
	and Kulish, Petr P.",
	title="New rational solutions of Yang-Baxter equation and deformed Yangians",
	journal="Czechoslovak Journal of Physics",
	year="1997",
	month="Jan",
	day="01",
	volume="47",
	number="1",
	pages="123--129",
	issn="1572-9486",
	doi="10.1023/A:1021460515598",
	url="https://doi.org/10.1023/A:1021460515598"
}

@article{Yu_2019,
	doi = {10.1088/1742-6596/1194/1/012117},
	url = {https://doi.org/10.1088\%2F1742-6596\%2F1194\%2F1\%2F012117},
		year = 2019,
		month = {apr},
		publisher = {{IOP} Publishing},
		volume = {1194},
		pages = {012117},
		author = {Li-Wei Yu and Mo-Lin Ge},
		title = {New type of solutions of Yang-Baxter equations, quantum entanglement and related physical models},
		journal = {Journal of Physics: Conference Series}
	}

@article{deLeeuw:2019zsi,
	author         = "de Leeuw, Marius and Pribytok, Anton and Ryan, Paul",
	title          = "{Classifying two-dimensional integrable spin chains}",
	year           = "2019",
	eprint         = "1904.12005",
	archivePrefix  = "arXiv",
	primaryClass   = "math-ph",
	SLACcitation   = "%%CITATION = ARXIV:1904.12005;%%"
}

@article{Vieira:2019vog,
	author         = "Vieira, R. S.",
	title          = "{Fifteen-vertex models with non-symmetric $R$ matrices}",
	year           = "2019",
	eprint         = "1908.06932",
	archivePrefix  = "arXiv",
	primaryClass   = "nlin.SI",
	SLACcitation   = "%%CITATION = ARXIV:1908.06932;%%"
}
@article{deLeeuw:2007akd,
      author         = "de Leeuw, M.",
      title          = "{Coordinate Bethe Ansatz for the String S-Matrix}",
      journal        = "J. Phys.",
      volume         = "A40",
      year           = "2007",
      pages          = "14413-14432",
      doi            = "10.1088/1751-8113/40/48/008",
      eprint         = "0705.2369",
      archivePrefix  = "arXiv",
      primaryClass   = "hep-th",
      reportNumber   = "ITP-UU-07-28, SPIN-07-18",
      SLACcitation   = "%%CITATION = ARXIV:0705.2369;%%"
}
@phdthesis{deLeeuw:2010nd,
      author         = "de Leeuw, Marius",
      title          = "{The S-matrix of the $AdS_5 x S^5$ superstring}",
      school         = "Utrecht U.",
      year           = "2010",
      eprint         = "1007.4931",
      archivePrefix  = "arXiv",
      primaryClass   = "hep-th",
      reportNumber   = "ITP-UU-10-13, SPIN-10-11",
      SLACcitation   = "%%CITATION = ARXIV:1007.4931;%%"
}
@article{Kulish:1985bj,
      author         = "Kulish, P. P.",
      title          = "{Integrable graded magnets}",
      journal        = "J. Sov. Math.",
      volume         = "35",
      year           = "1986",
      pages          = "2648-2662",
      doi            = "10.1007/BF01083770",
      note           = "[Zap. Nauchn. Semin.145,140(1985)]",
      SLACcitation   = "%%CITATION = JOSMA,35,2648;%%"
}
@article{Bracken:1994hz,
      author         = "Bracken, Anthony J. and Gould, Mark D. and Zhang,
                        Yao-Zhong and Delius, Gustav W.",
      title          = "{Solutions of the quantum Yang-Baxter equation with extra
                        nonadditive parameters}",
      journal        = "J. Phys.",
      volume         = "A27",
      year           = "1994",
      pages          = "6551-6562",
      doi            = "10.1088/0305-4470/27/19/025",
      eprint         = "hep-th/9405138",
      archivePrefix  = "arXiv",
      primaryClass   = "hep-th",
      reportNumber   = "UQMATH-94-03, BI-TP-94-23",
      SLACcitation   = "%%CITATION = HEP-TH/9405138;%%"
}
@article{batchelor2008quantum,
  title={The quantum inverse scattering method with anyonic grading},
  author={Batchelor, M and Foerster, A and Guan, X-W and Links, J and Zhou, H-Q},
  journal={Journal of physics. A, Mathematical and theoretical},
  volume={41},
  number={46},
  year={2008},
  publisher={IOP}
}

@article{Loebbert_2016,
	doi = {10.1088/1751-8113/49/32/323002},
	url = {https://doi.org/10.1088\%2F1751-8113\%2F49\%2F32\%2F323002},
	year = 2016,
	month = {jul},
	publisher = {{IOP} Publishing},
	volume = {49},
	number = {32},
	pages = {323002},
	author = {Florian Loebbert},
	title = {Lectures on Yangian symmetry},
	journal = {Journal of Physics A: Mathematical and Theoretical},
	abstract = {In these introductory lectures we discuss the topic of Yangian symmetry from various perspectives. Forming the classical counterpart of the Yangian and an extension of ordinary Noether symmetries, first the concept of nonlocal charges in classical, two-dimensional field theory is reviewed. We then define the Yangian algebra following Drinfel’d's original motivation to construct solutions to the quantum Yang–Baxter equation. Different realizations of the Yangian and its mathematical role as a Hopf algebra and quantum group are discussed. We demonstrate how the Yangian algebra is implemented in quantum, two-dimensional field theories and how its generators are renormalized. Implications of Yangian symmetry on the two-dimensional scattering matrix are investigated. We furthermore consider the important case of discrete Yangian symmetry realized on integrable spin chains. Finally we give a brief introduction to Yangian symmetry in planar, four-dimensional super Yang–Mills theory and indicate its impact on the dilatation operator and tree-level scattering amplitudes. These lectures are illustrated by several examples, in particular the two-dimensional chiral Gross–Neveu model, the Heisenberg spin chain and  superconformal Yang–Mills theory in four dimensions.}
}

@article{Essler:1991wg,
      author         = "Essler, Fabian H. L. and Korepin, Vladimir E. and
                        Schoutens, Kareljan",
      title          = "{Completeness of the SO(4) extended Bethe ansatz for the
                        one-dimensional Hubbard model}",
      journal        = "Nucl. Phys.",
      volume         = "B384",
      year           = "1992",
      pages          = "431-458",
      doi            = "10.1016/0550-3213(92)90575-V",
      eprint         = "cond-mat/9209012",
      archivePrefix  = "arXiv",
      primaryClass   = "cond-mat",
      reportNumber   = "ITP-SB-91-51",
      SLACcitation   = "%%CITATION = COND-MAT/9209012;%%"
}

@article{article,
author = {Heilmann, Ole and Lieb, Elliott},
year = {2006},
month = {12},
pages = {584 - 617},
title = {Violation of the Noncrossing Rule: The Hubbard Hamiltonian for Benzene},
volume = {172},
journal = {Annals of the New York Academy of Sciences},
doi = {10.1111/j.1749-6632.1971.tb34956.x}
}

@article{Beisert:2006qh,
      author         = "Beisert, Niklas",
      title          = "{The Analytic Bethe Ansatz for a Chain with Centrally
                        Extended su(2|2) Symmetry}",
      journal        = "J. Stat. Mech.",
      volume         = "0701",
      year           = "2007",
      pages          = "P01017",
      doi            = "10.1088/1742-5468/2007/01/P01017",
      eprint         = "nlin/0610017",
      archivePrefix  = "arXiv",
      primaryClass   = "nlin.SI",
      reportNumber   = "AEI-2006-074, PUTP-2211",
      SLACcitation   = "%%CITATION = NLIN/0610017;%%"
}
@article{hubbard1963electron,
  title={Electron correlations in narrow energy bands},
  author={Hubbard, John},
  journal={Proceedings of the Royal Society of London. Series A. Mathematical and Physical Sciences},
  volume={276},
  number={1365},
  pages={238--257},
  year={1963},
  publisher={The Royal Society London}
}

@article{Lieb:1968zza,
      author         = "Lieb, Elliott H. and Wu, F. Y.",
      title          = "{Absence of Mott transition in an exact solution of the
                        short-range, one-band model in one dimension}",
      journal        = "Phys. Rev. Lett.",
      volume         = "20",
      year           = "1968",
      pages          = "1445-1448",
      doi            = "10.1103/PhysRevLett.21.192.2,
                        10.1103/PhysRevLett.20.1445",
      note           = "[Erratum: Phys. Rev. Lett.21,192(1968)]",
      SLACcitation   = "%%CITATION = PRLTA,20,1445;%%"
}
@Article{SriramShastry1988,
author="Sriram Shastry, B.",
title="Decorated star-triangle relations and exact integrability of the one-dimensional Hubbard model",
journal="Journal of Statistical Physics",
year="1988",
month="Jan",
day="01",
volume="50",
number="1",
pages="57--79",
abstract="The exact integrability of the one-dimensional Hubbard model is demonstrated with the help of a novel set of triangle relations, the decorated star-triangle relations. The covering two-dimensional statistical mechanical model obeys the star-triangle or Yang-Baxter relation. A conjecture is presented for the eigenvalues of the transfer matrix.",
issn="1572-9613",
doi="10.1007/BF01022987",
url="https://doi.org/10.1007/BF01022987"
}

@article{KAROWSKI1979244,
title = "On the bound state problem in 1+1 dimensional field theories",
journal = "Nuclear Physics B",
volume = "153",
pages = "244 - 252",
year = "1979",
issn = "0550-3213",
doi = "https://doi.org/10.1016/0550-3213(79)90600-X",
url = "http://www.sciencedirect.com/science/article/pii/055032137990600X",
author = "M. Karowski",
abstract = "In the framework of factorizing S-matrices in 1+1 dimensions, further restrictions for the construction of S-matrices are discussed. A relation between residues of S-matrix poles and the parities of corresponding bound states is derived."
}

@article{Li:2018xrb,
      author         = "Li, Guang-Liang and Cao, Junpeng and Xue, Panpan and Xin,
                        Zhi-Rong and Hao, Kun and Yang, Wen-Li and Shi, Kangjie
                        and Wang, Yupeng",
      title          = "{Exact solution of the $sp(4)$ integrable spin chain with
                        generic boundaries}",
      journal        = "JHEP",
      volume         = "05",
      year           = "2019",
      pages          = "067",
      doi            = "10.1007/JHEP05(2019)067",
      eprint         = "1812.03618",
      archivePrefix  = "arXiv",
      primaryClass   = "math-ph",
      SLACcitation   = "%%CITATION = ARXIV:1812.03618;%%"
}

@article{Frolov:2011wg,
      author         = "Frolov, Sergey and Quinn, Eoin",
      title          = "{Hubbard-Shastry lattice models}",
      journal        = "J. Phys.",
      volume         = "A45",
      year           = "2012",
      pages          = "095004",
      doi            = "10.1088/1751-8113/45/9/095004",
      eprint         = "1111.5304",
      archivePrefix  = "arXiv",
      primaryClass   = "cond-mat.str-el",
      reportNumber   = "TCD-MATH-11-14, HMI-11-05",
      SLACcitation   = "%%CITATION = ARXIV:1111.5304;%%"
}

@article{Essler:1992py,
      author         = "Essler, Fabian H. L. and Korepin, Vladimir E. and
                        Schoutens, Kareljan",
      title          = "{New exactly solvable model of strongly correlated
                        electrons motivated by high T(c) superconductivity}",
      journal        = "Phys. Rev. Lett.",
      volume         = "68",
      year           = "1992",
      pages          = "2960-2963",
      doi            = "10.1103/PhysRevLett.68.2960",
      eprint         = "cond-mat/9209002",
      archivePrefix  = "arXiv",
      primaryClass   = "cond-mat",
      reportNumber   = "ITP-SB-92-03",
      SLACcitation   = "%%CITATION = COND-MAT/9209002;%%"
}

@article{Martins:2007hb,
      author         = "Martins, M. J. and Melo, C. S.",
      title          = "{The Bethe ansatz approach for factorizable centrally
                        extended S-matrices}",
      journal        = "Nucl. Phys.",
      volume         = "B785",
      year           = "2007",
      pages          = "246-262",
      doi            = "10.1016/j.nuclphysb.2007.05.021",
      eprint         = "hep-th/0703086",
      archivePrefix  = "arXiv",
      primaryClass   = "hep-th",
      reportNumber   = "UFSCAR-TH-07-03",
      SLACcitation   = "%%CITATION = HEP-TH/0703086;%%"
}
@inproceedings{Drummond:2007sa,
      author         = "Drummond, James and Feverati, Giovanni and Frappat, Luc
                        and Ragoucy, Eric",
      title          = "{Generalised integrable Hubbard models}",
      booktitle      = "{International Workshop on Recent Advances in Quantum
                        Integrable Systems (RAQIS 07) Annecy-le-Vieux, France,
                        September 11-14, 2007}",
      year           = "2007",
      eprint         = "0712.1940",
      archivePrefix  = "arXiv",
      primaryClass   = "hep-th",
      reportNumber   = "LAPTH-CONF-1222-07",
      SLACcitation   = "%%CITATION = ARXIV:0712.1940;%%"
}
@article{Drummond:2007gt,
      author         = "Drummond, J. M. and Feverati, G. and Frappat, L. and
                        Ragoucy, E.",
      title          = "{Super-Hubbard models and applications}",
      journal        = "JHEP",
      volume         = "05",
      year           = "2007",
      pages          = "008",
      doi            = "10.1088/1126-6708/2007/05/008",
      eprint         = "hep-th/0703078",
      archivePrefix  = "arXiv",
      primaryClass   = "hep-th",
      reportNumber   = "LAPTH-1176-07",
      SLACcitation   = "%%CITATION = HEP-TH/0703078;%%"
}

@ARTICLE{1998PhLA..239..187M,
       author = {{Maassarani}, Z.},
        title = "{The su( n) Hubbard model}",
      journal = {Physics Letters A},
     keywords = {HUBBARD MODEL, SU(N) SPIN CHAIN, INTEGRABILITY, Condensed Matter - Statistical Mechanics, High Energy Physics - Theory, Nonlinear Sciences - Exactly Solvable and Integrable Systems},
         year = 1998,
        month = mar,
       volume = {239},
       number = {3},
        pages = {187-190},
          doi = {10.1016/S0375-9601(97)00977-8},
archivePrefix = {arXiv},
       eprint = {cond-mat/9709252},
 primaryClass = {cond-mat.stat-mech},
       adsurl = {https://ui.adsabs.harvard.edu/abs/1998PhLA..239..187M},
      adsnote = {Provided by the SAO/NASA Astrophysics Data System}
}

@ARTICLE{1998NuPhB.517..395M,
       author = {{Maassarani}, Z. and {Mathieu}, P.},
        title = "{The su ( N) XX model}",
      journal = {Nuclear Physics B},
     keywords = {Condensed Matter - Statistical Mechanics, High Energy Physics - Theory, Mathematics - Quantum Algebra, Nonlinear Sciences - Exactly Solvable and Integrable Systems},
         year = 1998,
        month = apr,
       volume = {517},
       number = {1},
        pages = {395-408},
          doi = {10.1016/S0550-3213(98)80004-7},
archivePrefix = {arXiv},
       eprint = {cond-mat/9709163},
 primaryClass = {cond-mat.stat-mech},
       adsurl = {https://ui.adsabs.harvard.edu/abs/1998NuPhB.517..395M},
      adsnote = {Provided by the SAO/NASA Astrophysics Data System}
}

@ARTICLE{1994JPhy1...4.1151I,
       author = {{Idzumi}, Makoto and {Tokihiro}, Tetsuji and {Arai}, Masao},
        title = "{Solvable nineteen-vertex models and quantum spin chains of spin one}",
      journal = {Journal de Physique I},
         year = 1994,
        month = aug,
       volume = {4},
       number = {8},
        pages = {1151-1159},
          doi = {10.1051/jp1:1994245},
       adsurl = {https://ui.adsabs.harvard.edu/abs/1994JPhy1...4.1151I},
      adsnote = {Provided by the SAO/NASA Astrophysics Data System}
}

@ARTICLE{Frappat1,
       author = {{Cramp{\'e}}, N. and {Frappat}, L. and {Ragoucy}, E.},
        title = "{Classification of three-state Hamiltonians solvable by the coordinate Bethe ansatz}",
      journal = {Journal of Physics A Mathematical General},
     keywords = {Mathematical Physics, High Energy Physics - Theory},
         year = 2013,
        month = oct,
       volume = {46},
       number = {40},
          eid = {405001},
        pages = {405001},
          doi = {10.1088/1751-8113/46/40/405001},
archivePrefix = {arXiv},
       eprint = {1306.6303},
 primaryClass = {math-ph},
       adsurl = {https://ui.adsabs.harvard.edu/abs/2013JPhA...46N5001C},
      adsnote = {Provided by the SAO/NASA Astrophysics Data System}
}

@ARTICLE{Frappat2,
       author = {{Fonseca}, T. and {Frappat}, L. and {Ragoucy}, E.},
        title = "{R matrices of three-state Hamiltonians solvable by coordinate Bethe ansatz}",
      journal = {Journal of Mathematical Physics},
     keywords = {Mathematical Physics, High Energy Physics - Theory},
         year = 2015,
        month = jan,
       volume = {56},
       number = {1},
          eid = {013503},
        pages = {013503},
          doi = {10.1063/1.4905893},
archivePrefix = {arXiv},
       eprint = {1406.3197},
 primaryClass = {math-ph},
       adsurl = {https://ui.adsabs.harvard.edu/abs/2015JMP....56a3503F},
      adsnote = {Provided by the SAO/NASA Astrophysics Data System}
}

@ARTICLE{Frappat3,
       author = {{Cramp{\'e}}, N. and {Frappat}, L. and {Ragoucy}, E. and {Vanicat}, M.},
        title = "{3-state Hamiltonians associated to solvable 33-vertex models}",
      journal = {Journal of Mathematical Physics},
     keywords = {Mathematical Physics},
         year = 2016,
        month = sep,
       volume = {57},
       number = {9},
          eid = {093504},
        pages = {093504},
          doi = {10.1063/1.4962920},
archivePrefix = {arXiv},
       eprint = {1509.07589},
 primaryClass = {math-ph},
       adsurl = {https://ui.adsabs.harvard.edu/abs/2016JMP....57i3504C},
      adsnote = {Provided by the SAO/NASA Astrophysics Data System}
}
@incollection{kulish1990solutions,
  title={Solutions of the Yang-Baxter equation},
  author={Kulish, PP and Sklyanin, EK},
  booktitle={Yang-Baxter Equation in Integrable Systems},
  pages={172--196},
  year={1990},
  publisher={World Scientific}
}

\end{bibtex}

\bibliographystyle{nb}
\bibliography{\jobname}

\end{document}